\documentclass[preprint,12pt]{elsarticle}

\usepackage{amssymb}
\usepackage{amsmath}
\usepackage{hyperref}
\usepackage{tikz,pgfplots}
\usepackage{todonotes}

\journal{Environmental Modelling \& Software}

\newcommand{\pig}{p_{\mathrm{ig}}}

\begin{document}

\begin{frontmatter}


\title{Statistical Models of Ember Wash and Their Impact on Wildfire
  Area Growth}

\author[label1,label2]{Bryan Quaife}
\ead{bquaife@fsu.edu}
\author[label1,label2]{Kevin Speer}
\affiliation[label1]{organization = {Department of Scientific Computing,
  Florida State University},
  addressline = {400 Dirac Science Library},
  city = {Tallahassee},
  postcode = {32306},
  state = {Florida},
  country = {USA}}
  \affiliation[label2]{organization = {Geophysical Fluid Dynamics
  Institute, Florida State University},
  addressline = {018 Keen Building},
  city = {Tallahassee},
  postcode = {32306},
  state = {Florida},
  country = {USA}}


\title{}


\author{} 


\begin{abstract}
Wildfire spread is strongly influenced by the transport and ignition of
  embers. While long-range spotting driven by plume lofting has received
  significant attention, embers transported near the surface by
  turbulent winds can also influence fire propagation. We develop a
  stochastic model for near-surface ember transport, referred to as {\em
  ember wash}. The model represents ember motion as a sequence of short
  displacements analogous to saltation-like transport and incorporates a
  probabilistic ignition process that depends on ember survival during
  transport. This formulation leads to an exponential distribution of
  ember flight times. The model is implemented within a simplified fire
  spread model to examine burn patterns and growth dynamics. Simulations
  demonstrate that ember wash produces spread behavior that differs
  fundamentally from classical plume-driven spotting. These results
  suggest that ember wash provides a plausible mechanism for wildfire
  spread regimes that differ from those predicted by geometric or
  plume-driven spread models.

\end{abstract}

%


\end{frontmatter}



\section{Introduction}
Wildland fires are complex, multiscale phenomena whose growth depends on
interactions among fuels, topography, atmospheric conditions, and
fire-atmosphere coupling. Predicting the evolution of wildland fires is
a central challenge for both operational forecasting and scientific
modeling. Wildland fire growth is commonly characterized in terms of
motion normal to the fire front, with a rate of spread that is
parameterized in terms of empirical evidence, physics, or data-driven
approaches. The rate of spread is a local measurement which determines
global behaviors including the total burned area, $A(t)$, and its rate
of growth, $dA/dt$. In practice, heterogeneous fuels, complex
landscapes, meteorological conditions, fire-atmosphere interactions, and
nonlocal transport processes collectively determine the rate of area
growth. Quantifying the roles of such terms provides a useful lens for
identifying the physical mechanisms that dominate fire spread under
different conditions. For example, it is natural to ask if this occurs
through the formation of multiple ignitions ahead of the main fire
caused by embers. In this work, we discuss processes that control the
rate of growth of the burn scar.

If a fire grows in all directions at a constant rate of spread, then the
radius of the burn scar will grow linearly with time, and the result is
a quadratic area growth scaling $A(t) \sim t^2$. This quadratic fire
area growth would indicate that the rate of growth $dA/dt$ is increasing
in time, that is, the fire is accelerating. However, this simple radial
growth is often nonphysical because fire-driven winds focus spread at
the head and suppress spread at the flanks and
rear~\cite{coe-cam-mic-pat-rig-yed2013}. Elliptical growth models are
one way to take this effect into account. Also, fires sometimes spread
more diffusively~\cite{beb-oli-qua-sko-hei-spe2020} leading to the
linear area growth scaling $A(t) \sim t$. Local conditions, such as
intermittent downdrafts or topographic slope, might transition the fire
growth between linear and quadratic scalings, but on the scale of the
whole fire, other processes are necessary for the fire area to produce
such rapid growth for extended periods of time. The central goal of this
paper is to reveal physical processes that control the transition
between linear and quadratic growth in a fire's area.

Spotting by embers lofted from the fire front are a well-known process
for extreme fire spread, and much effort has gone into the study of the
mechanisms involved in the different components of spotting
(e.g.~\cite{albini2012mathematical}). Additionally, at the wildland
urban interface (WUI), embers are responsible for destruction of large
urban areas such as the Marshall Fire in Boulder County, Colorado, and
the Camp Fire that destroyed Paradise, California. Models have
demonstrated the important role of plume intensity and turbulence in the
degree and range of spotting, and suggested ``short-range''
exponential-like distributions of spotting downwind of the
fire~\cite{martin2016spotting, bhutia2010comparison}. Data analyses
confirm some aspects of this short-range behavior in diverse fuel and
wind conditions based on fires in Australia~\cite{storey2021experiments}
and the Rocky Mountains, USA~\cite{page2019analysis}.

\begin{figure}[htp]
  \centering
  \includegraphics[width=0.9\textwidth]{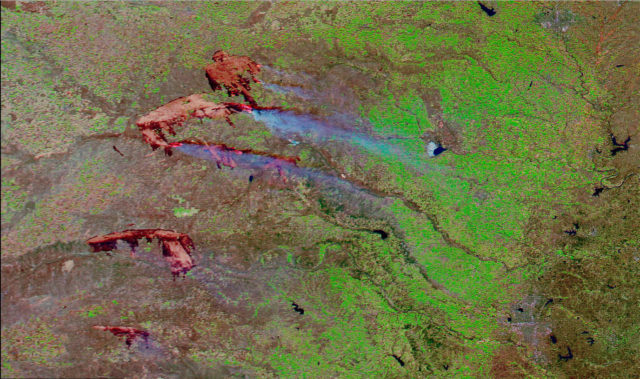}
  \caption{\label{fig:OK2} \em Infrared imagery of the 2017 NW OK
  Complex Fire near Woodward showing the intense fingering and complex
  spread behavior, associated with ember transport, wind and fuel
  variations, and river drainage slopes. The Complex Fire consisted of
  several major fires and burned over 834,000~acres. Wind speeds were
  over 30~mph (15~m/s) with gusts up to 50~mph (25~m/s) in places. Flame
  heights were reported in places at over 40~ft. Credit: OK Forest
  Service Public Information Map.} 
\end{figure}

\begin{figure}[htp]
\centering
  \includegraphics[width=1.0\textwidth]{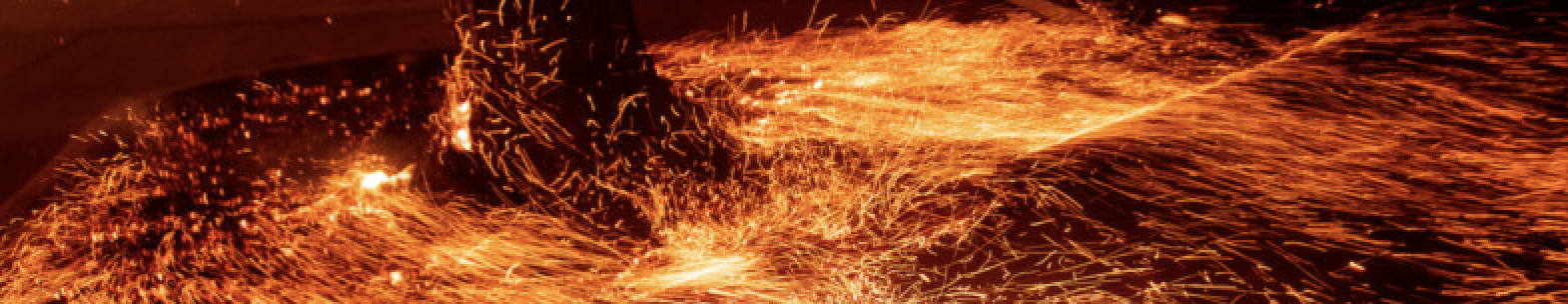}
  \caption{\label{fig:EmbersTree} \em Embers transported along the
  ground in the surface mode of fire spread. Tree and roughness element
  wake interactions, particle hopping, and jumping effects are
  illustrated. Credit: Pexel/PublicDomain.} 
\end{figure}

Short-range spotting appears to be an integral part of fire front
movement in higher winds across most fuel types, including natural and
man-made. An example of rapid spread thought to involve short-range
ember transport appears in the Northwest Oklahoma Fire Complex of three
large wildfires that burned almost 780,000~acres in Oklahoma and Kansas
(Figure~\ref{fig:OK2}). Some of this short range ``spotting'' is due to
firebrands that did not rise in the fire's plume but rather spread
nearly horizontally in the surface boundary layer
(Figure~\ref{fig:EmbersTree}). These and the heavier embers and burning
debris that emerge from the main fire frontal region carried by strong
winds are responsible for rapid movement and widening of the front. We
use the term ``ember wash'' to denote this component of wildland fire
spread.

It is helpful to first more precisely compare ember wash and standard
spotting phenomenon. Both ember wash and spotting involve the generation
and release of embers into the fire environment. This occurs due to
sources from natural vegetation and from man-made structures, such as a
residential housing, commercial buildings, vehicles, and fences. Once
released, the embers closest to the fire may be entrained or directly
injected into the buoyant plume and experience rapid lofting, which is a
fundamental aspect of spotting. A great deal of effort has gone into a
better understanding of the detailed mechanisms and mechanical forces
that are applied to embers during this process~\cite{sardoy2007modeling,
albini2012mathematical, tohidi2017stochastic}. Along with these aspects,
the combustion characteristics of embers in flight, shape, mechanical
forces and stresses, composition, and mass loss are important to
consider for their ultimate role in igniting
structures~\cite{manzello2020role, dos-yag2023, wad-sut-ooi-moi2022}.
Koo {\em et al.} provide an excellent overview of spotting ignition
processes~\cite{koo-pag-wei-woy2010}.

Ember wash by contrast is fundamentally a surface boundary layer process
in which embers are released and remain on or close to the ground. These
embers are influenced not only by plume processes but also by
near-surface turbulent structures~\cite{des-goo-ban2022}. The embers may
be generated outside the main fire and ignite along their path, and they
may leave and re-enter the fire front or pass through other fires as
they proceed. Embers are released and resuspended by energetic,
turbulent ground-level winds which largely depend on the arrangement of
natural and man-made structures. Hence, the ember wash process is more
akin to sediment transport than to inertial particle transport in the
atmosphere, although the latter forms one component near the ground too.
More colorful descriptions include ember storm or blizzard, the latter
appropriate as the movement is mostly horizontal; however, we should
point out that ember wash as defined here is not the same as the
sediment ``wash load'' which consists of the fine suspended particle
component only and does not interact with the bottom boundary
significantly~\cite{fur-sch-sch-fat2016}. 

Real fires spread rapidly in extreme conditions by the strong
wind-driven front, spotting due to lofted embers and those traveling
along the surface. In modeling terms, the key difference between lofted
and surface embers is that while embers lofted in the buoyant plume are
subject to transport in the higher level background wind (above the
canopy or local terrain), the embers transported near the ground are
carried by local surface winds and near-boundary layer turbulence.
Generally, the ember wash occurs outside the surface frictional layer,
driven by flow that varies at scales comparable to the roughness
elements of the boundary and shear layer eddies. The presence of a
canopy and structures will introduce its own eddy turbulence scales and
horizontal mixing~\cite{beb-oli-qua-sko-hei-spe2020}. Ejections of
burning embers to higher levels by boundary layer turbulence and eddies
could also give rise to what would effectively be a mode of surface
ember transport. Clearly, various transport processes may all be
operating at once in the complex 3D turbulent flow near the surface. 

Our objective is to represent in an idealized fashion some of the most
important surface processes for wind-driven fire in a turbulent flow. We
show below how complex fire fronts and effects emerge from the
interaction of background wind, fire-induced flow, and a new ember wash
effect near intense fires. We develop a statistical model for ember wash
based on analogies with sediment transport and survival processes, and
couple it to an idealized fire spread model to examine transitions
between linear, intermediate, and quadratic growth regimes. Ultimately,
we show that a near-surface stochastic ember transport mechanism is
sufficient to generate transitions between linear and quadratic
fire-area growth.

The remainder of the paper is organized as follows. In
Section~\ref{sec:GIS}, we use remotely sensed observations to
demonstrate that fire growth can be both linear and quadratic. In
Section~\ref{sec:saltation}, we propose a new model for ember wash that
runs in parallel to saltation models used to model sand transport.
Sections~\ref{sec:survival} and~\ref{sec:fire_model} summarize a
simplified fire spread model and couples it with ember dynamics with
statistical concepts from survival analysis. Finally,
Sections~\ref{sec:results} and~\ref{sec:discussion} show results from
simulations and discuss fundamental differences between short- and
long-range ember transport.

\section{GIS-Derived Fire Area Growth and Scaling Behavior}
\label{sec:GIS}

\begin{figure}[htp]
  \centering
  \scalebox{1}{\input{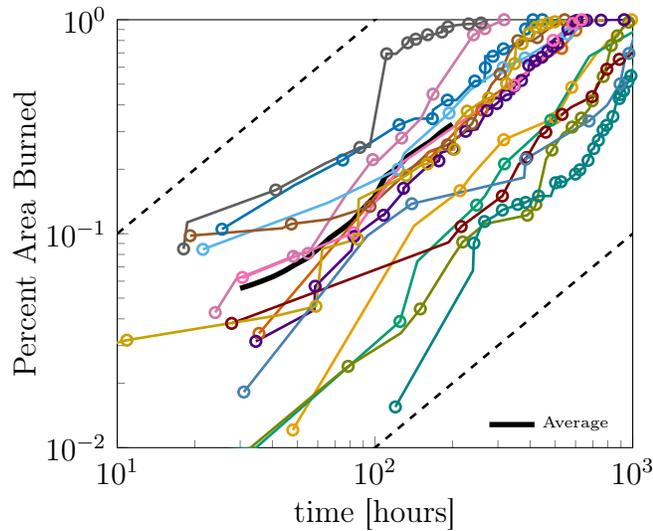}}
  \caption{\label{fig:historical1} \em The burnt area of 15 wildfires in
  the Western USA. All areas are normalized according to their final
  size. The slope of the dashed black lines is 1, indicating a linear
  growth in the fire's area.}
\end{figure}

\begin{figure}[htp]
  \begin{center}
    \fbox{\includegraphics[width=0.225\textwidth,trim=7cm 7cm
    7cm 7cm,clip]{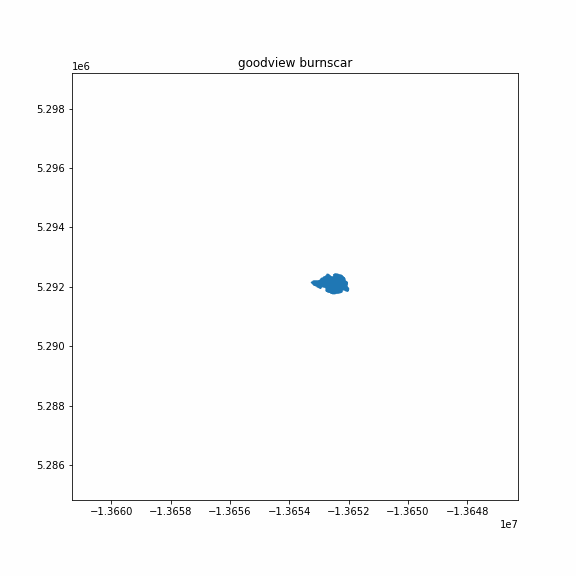}}
    \fbox{\includegraphics[width=0.225\textwidth,trim=7cm 7cm
    7cm 7cm,clip]{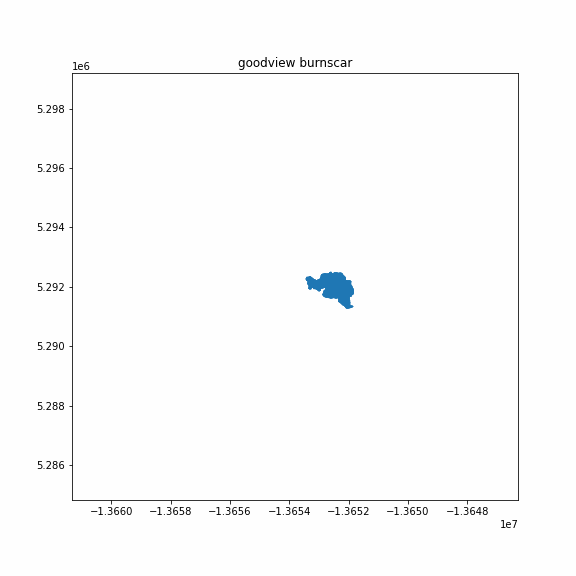}}
    \fbox{\includegraphics[width=0.225\textwidth,trim=7cm 7cm
    7cm 7cm,clip]{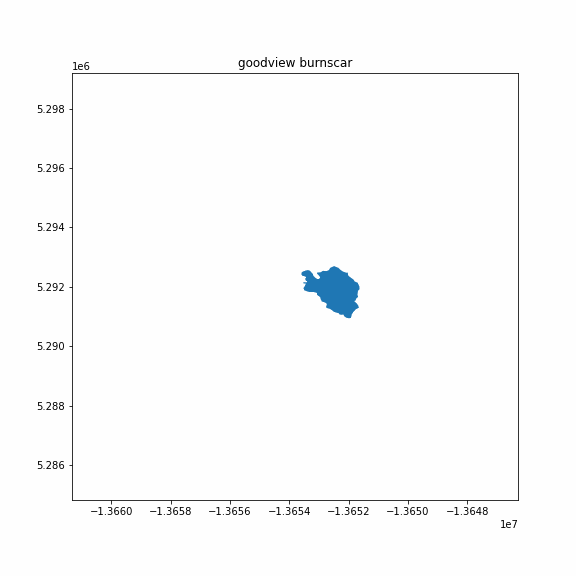}}
    \fbox{\includegraphics[width=0.225\textwidth,trim=7cm 7cm
    7cm 7cm,clip]{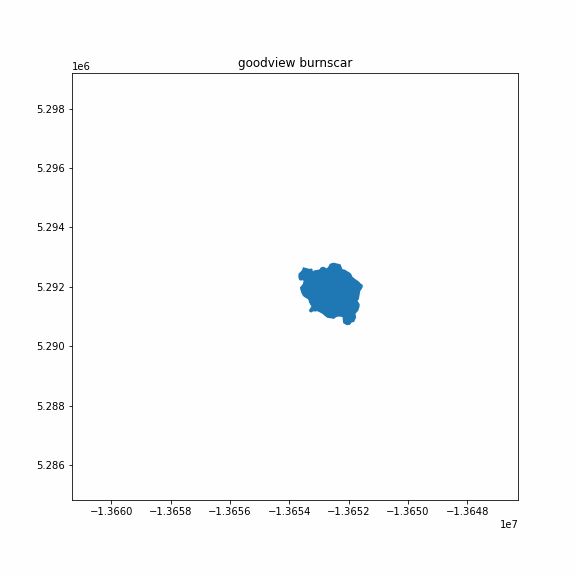}} \\[2pt]
    \fbox{\includegraphics[width=0.225\textwidth,trim=7cm 7cm
    7cm 7cm,clip]{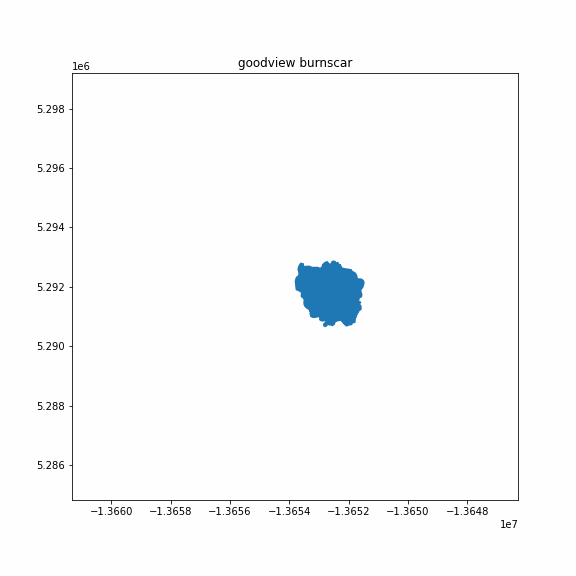}}
    \fbox{\includegraphics[width=0.225\textwidth,trim=7cm 7cm
    7cm 7cm,clip]{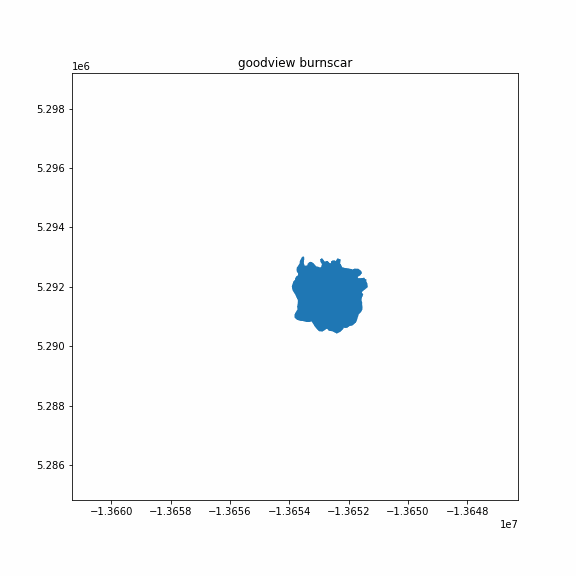}}
    \fbox{\includegraphics[width=0.225\textwidth,trim=7cm 7cm
    7cm 7cm,clip]{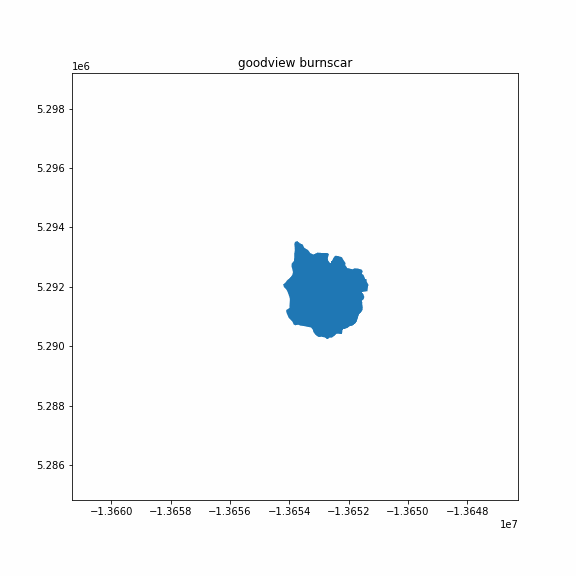}}
    \fbox{\includegraphics[width=0.225\textwidth,trim=7cm 7cm
    7cm 7cm,clip]{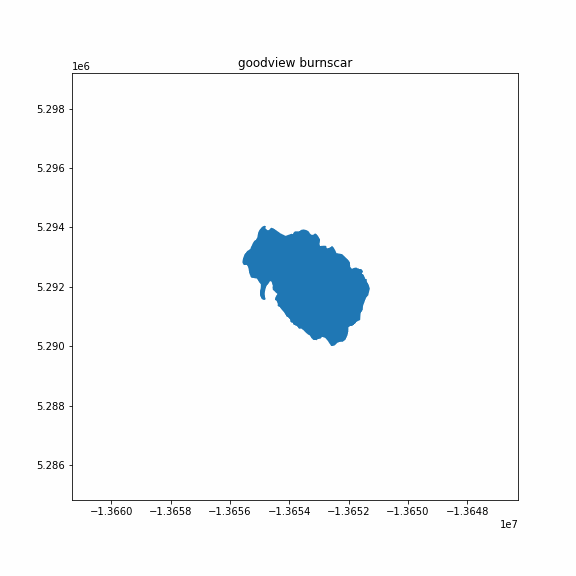}} \\[2pt]
    \fbox{\includegraphics[width=0.225\textwidth,trim=7cm 7cm
    7cm 7cm,clip]{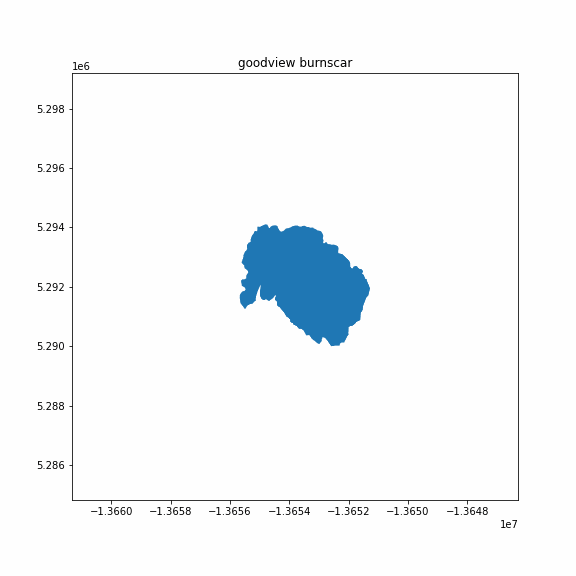}}
    \fbox{\includegraphics[width=0.225\textwidth,trim=7cm 7cm
    7cm 7cm,clip]{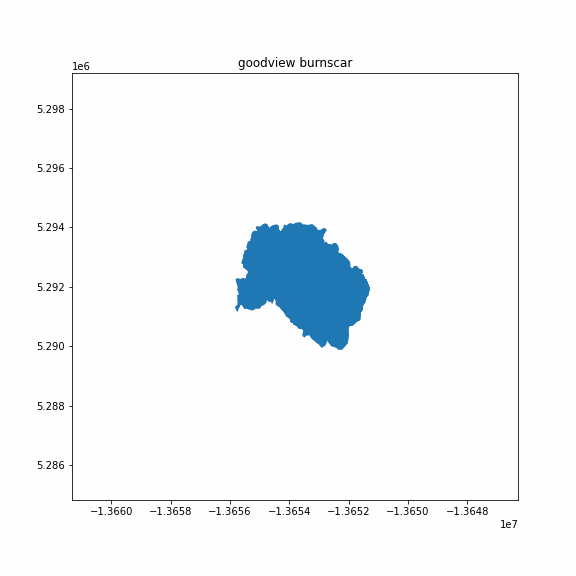}}
    \fbox{\includegraphics[width=0.225\textwidth,trim=7cm 7cm
    7cm 7cm,clip]{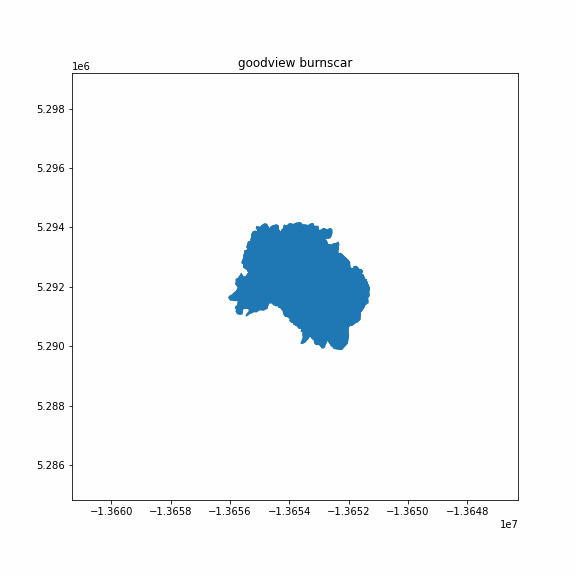}}
    \fbox{\includegraphics[width=0.225\textwidth,trim=7cm 7cm
    7cm 7cm,clip]{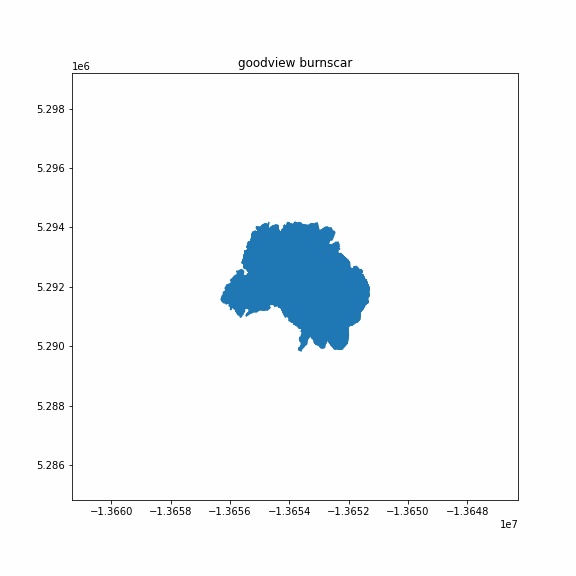}}
  \end{center}
  \caption{\label{fig:goodview} \em The Goodview fire exhibits the
  linear area scaling $A(t) \sim t$. Notice the lack of spotting.}
\end{figure}

To investigate the asymptotic growth of burn scars from historical
fires, we process GIS data of burn scars of over 500 fires that occurred
in the western USA between 2015 and 2021. Many are poorly sampled in the
growth period, or reach suppression-based or other physical limits after
relatively few samples. Of the burn scar data, 22 fire events have
enough samples to be able to characterize significant asymptotic trends
in the area growth.
Figure~\ref{fig:historical1} shows the area as a function of time of 15
wildland fires from the western USA. The black dashed lines have slope
1, indicating that the area of these fires is growing linearly with
time. The solid black line is an average over these 15 fire events. One
typical fire event in Figure~\ref{fig:historical1} is the Goodview fire,
and GIS frames of the burn scar at different times are provided in
Figure~\ref{fig:goodview}. One characteristic that we can observe is the
regular spreading and lack of spotting.

\begin{figure}[htp]
  \begin{center}
    \scalebox{1}{
\definecolor{OIblue}{RGB}{0,114,178}
\definecolor{OIorange}{RGB}{230,159,0}
\definecolor{OIgreen}{RGB}{0,158,115}
\definecolor{OIvermillion}{RGB}{213,94,0}
\definecolor{OIpurple}{RGB}{204,121,167}
\definecolor{OIcyan}{RGB}{86,180,233}
\definecolor{OIgray}{RGB}{94,94,94}

\begin{tikzpicture}[scale=1]

\begin{axis}[
  /pgfplots/tick scale binop=\times,
  xmode = log,
  ymode = log,
  xmin = 8e0,
  xmax = 1.2e3,
  ymin = 1e-3,
  ymax = 1,
  xtick = {1e1,1e2,1e3},
  ytick = {1e-3, 1e-2, 1e-1, 1e0},
  xlabel = {\normalsize time [hours]},
  ylabel = {Percent Area Burned},
  xtick pos = left,
  ytick pos = left,
  legend entries = {Tepee Springs,Cove,Haypress,Nu
  Copeland,Schaeffer,Springs,Wallow,Average},
  legend cell align=left,
  legend style={draw=none,font=\tiny,fill=none},
  legend style={at={(0.98,0.01)},anchor=south east} 
]

\addplot[
  color=OIblue,
  line width=1.0pt,
  mark=o,mark repeat=2] coordinates{
(0.00e+00,7.75e-03)
(2.40e+01,1.99e-02)
(4.85e+01,4.50e-02)
(7.43e+01,4.83e-02)
(9.19e+01,5.15e-02)
(9.60e+01,6.16e-02)
(1.18e+02,6.17e-02)
(1.45e+02,6.85e-02)
(1.68e+02,7.21e-02)
(1.94e+02,8.53e-02)
(2.12e+02,1.02e-01)
(2.19e+02,1.03e-01)
(2.40e+02,1.14e-01)
(2.40e+02,1.14e-01)
(2.51e+02,1.14e-01)
(2.66e+02,1.30e-01)
(2.73e+02,1.31e-01)
(2.85e+02,1.55e-01)
(3.09e+02,1.90e-01)
(3.24e+02,1.96e-01)
(3.32e+02,2.68e-01)
(3.42e+02,2.68e-01)
(3.58e+02,4.18e-01)
(3.72e+02,4.95e-01)
(3.81e+02,5.04e-01)
(3.99e+02,7.95e-01)
(4.06e+02,8.12e-01)
(4.19e+02,8.55e-01)
(4.29e+02,8.86e-01)
(4.32e+02,9.39e-01)
(4.53e+02,9.46e-01)
(4.73e+02,9.60e-01)
(4.78e+02,9.71e-01)
(4.95e+02,9.74e-01)
(5.24e+02,9.89e-01)
(5.35e+02,9.89e-01)
(5.63e+02,9.89e-01)
(5.77e+02,9.89e-01)
(6.21e+02,9.90e-01)
(6.45e+02,9.90e-01)
(6.47e+02,9.90e-01)
(6.54e+02,9.90e-01)
(6.70e+02,9.92e-01)
(6.94e+02,9.93e-01)
(7.01e+02,9.93e-01)
(7.16e+02,9.96e-01)
(7.40e+02,9.97e-01)
(7.53e+02,9.97e-01)
(7.69e+02,1.00e+00)
(7.77e+02,1.00e+00)
(7.87e+02,1.00e+00)
(8.02e+02,1.00e+00)
(8.36e+02,1.00e+00)
(8.61e+02,1.00e+00)
(9.33e+02,1.00e+00)
(9.33e+02,1.00e+00)
(9.56e+02,1.00e+00)
(9.81e+02,1.00e+00)
(9.86e+02,1.00e+00)
(1.05e+03,1.00e+00)
};

\addplot[
  color=OIorange,
  line width=1pt,
  mark=o,mark repeat=2] coordinates{
(0.00e+00,2.75e-03)
(2.17e+01,5.16e-03)
(2.71e+01,5.16e-03)
(5.06e+01,6.27e-02)
(6.95e+01,1.05e-01)
(9.41e+01,2.20e-01)
(1.18e+02,3.67e-01)
(1.18e+02,3.67e-01)
(1.46e+02,7.29e-01)
(1.65e+02,9.24e-01)
(1.67e+02,9.60e-01)
(1.91e+02,9.97e-01)
(2.15e+02,1.00e+00)
(2.66e+02,1.00e+00)
(2.93e+02,1.00e+00)
(3.11e+02,1.00e+00)
(3.59e+02,1.00e+00)
(4.11e+02,1.00e+00)
};

\addplot[
  color=OIgreen,
  line width=1pt,
  mark=o,mark repeat=2] coordinates{
(0.00e+00,7.66e-04)
(4.68e+01,6.03e-03)
(4.99e+01,1.45e-02)
(5.99e+01,1.72e-02)
(9.44e+01,3.65e-02)
(9.74e+01,3.78e-02)
(1.20e+02,3.78e-02)
(1.30e+02,8.48e-02)
(1.46e+02,8.65e-02)
(1.92e+02,1.19e-01)
(2.16e+02,1.27e-01)
(2.17e+02,1.27e-01)
(2.40e+02,1.27e-01)
(2.45e+02,1.41e-01)
(2.64e+02,1.67e-01)
(2.65e+02,1.70e-01)
(2.93e+02,1.94e-01)
(3.16e+02,2.03e-01)
(3.40e+02,2.16e-01)
(3.60e+02,2.28e-01)
(3.68e+02,2.28e-01)
(3.89e+02,2.53e-01)
(4.16e+02,2.86e-01)
(4.41e+02,3.65e-01)
(4.55e+02,3.95e-01)
(4.89e+02,4.48e-01)
(5.04e+02,5.24e-01)
(5.12e+02,5.24e-01)
(5.57e+02,5.40e-01)
(5.81e+02,5.97e-01)
(6.06e+02,6.37e-01)
(6.28e+02,7.95e-01)
(6.47e+02,9.47e-01)
(6.73e+02,9.47e-01)
(6.97e+02,9.47e-01)
(7.22e+02,9.53e-01)
(7.45e+02,9.58e-01)
(7.46e+02,9.63e-01)
(7.69e+02,9.63e-01)
(7.94e+02,9.65e-01)
(8.17e+02,9.75e-01)
(8.18e+02,9.82e-01)
(8.42e+02,9.86e-01)
(8.64e+02,9.86e-01)
(8.88e+02,9.86e-01)
(9.12e+02,9.86e-01)
(9.37e+02,9.88e-01)
(9.58e+02,9.88e-01)
(9.85e+02,9.99e-01)
(9.86e+02,1.00e+00)
(1.03e+03,1.00e+00)
(1.06e+03,1.00e+00)
};

\addplot[
  color=OIvermillion,
  line width=1pt,
  mark=o,mark repeat=2] coordinates{
(0.00e+00,5.09e-03)
(2.51e+01,5.09e-03)
(4.91e+01,9.75e-03)
(7.69e+01,9.91e-03)
(9.72e+01,2.10e-02)
(1.22e+02,5.88e-02)
(1.43e+02,6.86e-02)
(1.69e+02,7.35e-02)
(1.90e+02,8.01e-02)
(3.10e+02,4.37e-01)
(3.13e+02,4.38e-01)
(3.58e+02,4.40e-01)
(4.08e+02,4.74e-01)
(4.56e+02,5.73e-01)
(4.78e+02,6.16e-01)
(5.06e+02,7.30e-01)
(5.27e+02,8.64e-01)
(5.51e+02,9.98e-01)
(9.27e+02,1.00e+00)
};

\addplot[
  color=OIpurple,
  line width=1pt,
  mark=o,mark repeat=2] coordinates{
(0.00e+00,1.23e-02)
(6.49e+01,5.07e-02)
(1.52e+02,2.79e-01)
(1.74e+02,3.08e-01)
(1.77e+02,3.08e-01)
(2.00e+02,5.22e-01)
(2.28e+02,5.31e-01)
(2.48e+02,7.13e-01)
(2.49e+02,7.13e-01)
(2.73e+02,7.14e-01)
(2.73e+02,7.56e-01)
(2.97e+02,8.81e-01)
(3.21e+02,9.17e-01)
(3.26e+02,9.18e-01)
(3.46e+02,9.41e-01)
(3.50e+02,9.69e-01)
(3.66e+02,9.76e-01)
(3.72e+02,9.76e-01)
(3.74e+02,9.76e-01)
(3.93e+02,9.76e-01)
(3.93e+02,9.81e-01)
(4.06e+02,9.81e-01)
(4.14e+02,9.98e-01)
(4.41e+02,9.98e-01)
(4.42e+02,1.00e+00)
(4.66e+02,1.00e+00)
(4.89e+02,1.00e+00)
(5.10e+02,1.00e+00)
};

\addplot[
  color=OIcyan,
  line width=1pt,
  mark=o,mark repeat=2] coordinates{
(0.00e+00,4.31e-03)
(2.26e+01,8.40e-03)
(4.96e+01,1.19e-02)
(1.02e+02,3.51e-02)
(1.19e+02,5.00e-02)
(1.45e+02,7.01e-02)
(1.67e+02,9.19e-02)
(1.97e+02,1.57e-01)
(2.14e+02,1.86e-01)
(2.43e+02,2.00e-01)
(2.65e+02,2.42e-01)
(2.69e+02,2.61e-01)
(2.91e+02,3.43e-01)
(3.12e+02,3.70e-01)
(3.40e+02,3.98e-01)
(3.63e+02,4.53e-01)
(3.76e+02,5.28e-01)
(3.86e+02,5.52e-01)
(4.10e+02,6.17e-01)
(4.41e+02,6.31e-01)
(4.92e+02,6.35e-01)
(5.23e+02,6.50e-01)
(5.56e+02,6.64e-01)
(6.05e+02,7.10e-01)
(6.29e+02,7.45e-01)
(6.50e+02,7.82e-01)
(6.72e+02,8.11e-01)
(6.91e+02,8.99e-01)
(7.19e+02,9.86e-01)
(7.34e+02,9.99e-01)
(7.56e+02,9.99e-01)
(7.86e+02,9.99e-01)
(7.99e+02,9.99e-01)
(8.68e+02,1.00e+00)
};

\addplot[
  color=OIgray,
  line width=1pt,
  mark=o,mark repeat=2] coordinates{
(0.00e+00,4.14e-03)
(8.98e+00,4.60e-03)
(7.44e+01,6.79e-03)
(8.55e+01,1.36e-02)
(1.06e+02,2.50e-02)
(1.35e+02,3.23e-02)
(1.63e+02,5.65e-02)
(1.82e+02,7.68e-02)
(1.83e+02,7.68e-02)
(2.09e+02,7.68e-02)
(2.11e+02,1.00e-01)
(2.29e+02,1.11e-01)
(2.59e+02,1.44e-01)
(2.77e+02,1.52e-01)
(3.07e+02,1.60e-01)
(3.29e+02,1.81e-01)
(3.58e+02,2.28e-01)
(3.76e+02,2.47e-01)
(4.01e+02,2.86e-01)
(4.24e+02,3.01e-01)
(4.48e+02,3.01e-01)
(4.50e+02,4.35e-01)
(4.74e+02,4.35e-01)
(4.99e+02,6.93e-01)
(5.17e+02,6.97e-01)
(5.20e+02,6.97e-01)
(5.44e+02,8.40e-01)
(5.69e+02,9.12e-01)
(5.87e+02,9.26e-01)
(5.89e+02,9.68e-01)
(5.91e+02,9.79e-01)
(5.92e+02,9.79e-01)
(6.17e+02,9.81e-01)
(6.40e+02,9.81e-01)
(6.66e+02,9.84e-01)
(7.13e+02,9.90e-01)
(7.33e+02,9.91e-01)
(7.58e+02,9.91e-01)
(7.59e+02,9.95e-01)
(7.82e+02,9.95e-01)
(7.84e+02,9.97e-01)
(8.04e+02,9.98e-01)
(8.29e+02,9.98e-01)
(8.54e+02,9.98e-01)
(8.78e+02,9.98e-01)
(9.10e+02,9.99e-01)
(9.26e+02,9.99e-01)
(9.28e+02,9.99e-01)
(9.73e+02,9.99e-01)
(9.98e+02,9.99e-01)
(1.02e+03,9.99e-01)
(1.02e+03,9.99e-01)
(1.07e+03,9.99e-01)
(1.08e+03,9.99e-01)
(1.12e+03,1.00e+00)
(1.19e+03,1.00e+00)
(1.31e+03,1.00e+00)
(1.38e+03,1.00e+00)
(1.46e+03,1.00e+00)
};

\addplot[
  color=black,
  line width=2pt] coordinates{
(2.00e+01,8.41e-03)
(2.31e+01,9.02e-03)
(2.66e+01,9.82e-03)
(3.07e+01,1.14e-02)
(3.54e+01,1.44e-02)
(4.08e+01,1.87e-02)
(4.70e+01,2.36e-02)
(5.42e+01,2.89e-02)
(6.25e+01,3.32e-02)
(7.21e+01,4.01e-02)
(8.32e+01,5.31e-02)
(9.59e+01,7.40e-02)
(1.11e+02,1.02e-01)
(1.28e+02,1.36e-01)
(1.47e+02,1.93e-01)
(1.70e+02,2.41e-01)
(1.96e+02,2.89e-01)
(2.26e+02,3.18e-01)
(2.60e+02,3.82e-01)
(3.00e+02,4.58e-01)
};

\addplot[
  color=black,dashed,
  line width=1pt] coordinates{
  (1e1,6e-3)
  (1e3,6e1)
};

\addplot[
  color=black,dashed,
  line width=1pt] coordinates{
  (1e1,1e-4)
  (1e3,1e0)
};

\end{axis}


\end{tikzpicture}}
  \end{center}
  \caption{\label{fig:historical2} \em The burnt area of seven wildfires
  in the western USA. All areas are normalized according to their final
  size. The slope of the black lines is 2, indicating a quadratic growth
  in the fire's area.}
\end{figure}
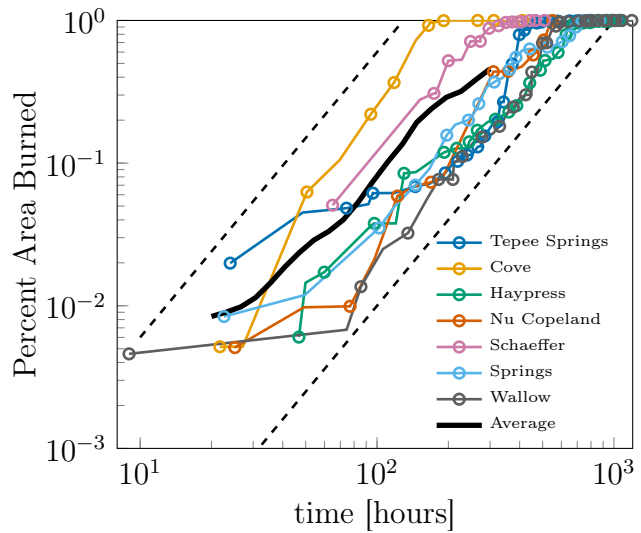

\begin{figure}[htp]
  \begin{center}
    \fbox{\includegraphics[width=0.225\textwidth,trim=5cm 5cm
    5cm 5cm,clip]{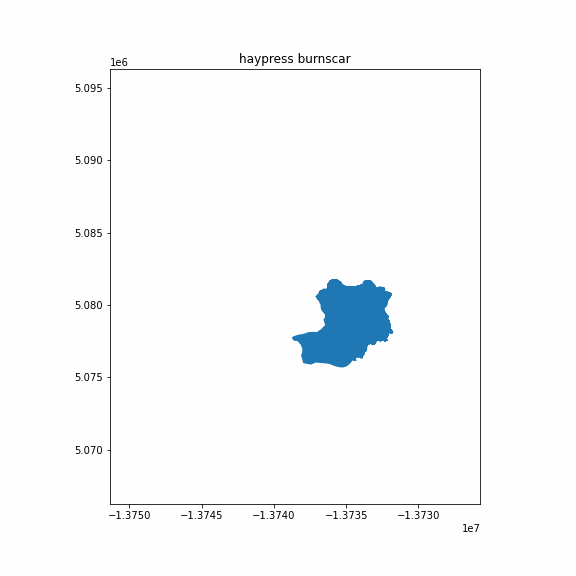}}
    \fbox{\includegraphics[width=0.225\textwidth,trim=5cm 5cm
    5cm 5cm,clip]{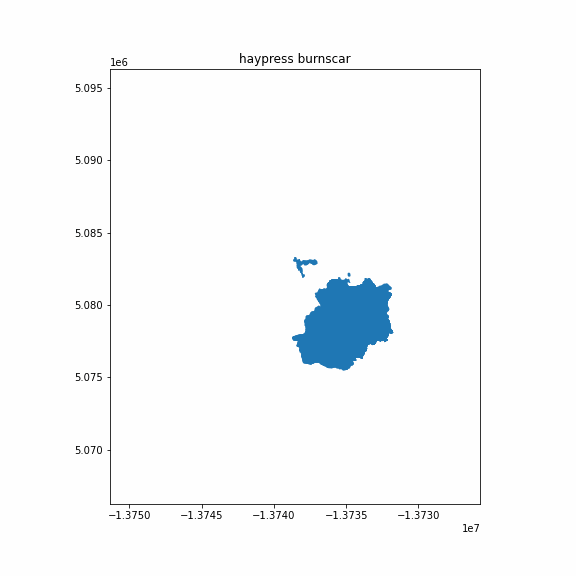}}
    \fbox{\includegraphics[width=0.225\textwidth,trim=5cm 5cm
    5cm 5cm,clip]{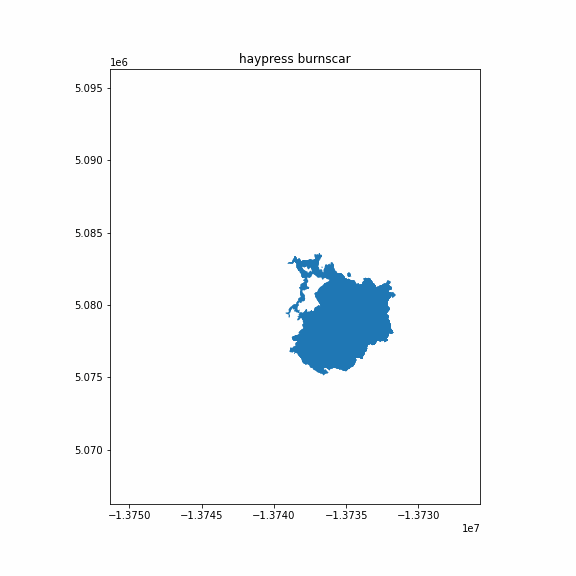}}
    \fbox{\includegraphics[width=0.225\textwidth,trim=5cm 5cm
    5cm 5cm,clip]{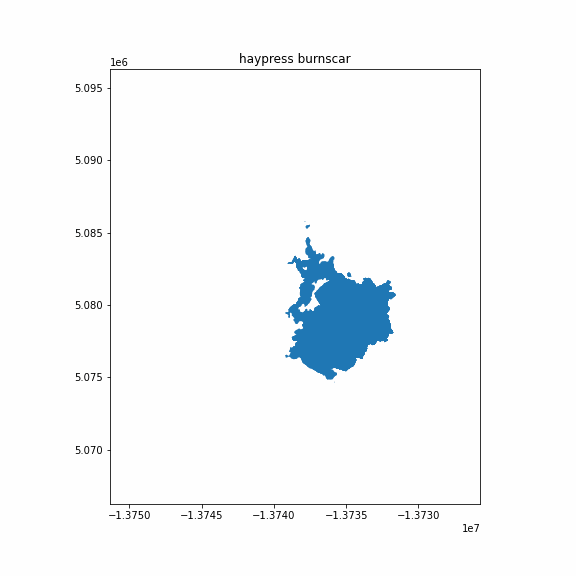}} \\[2pt]
    \fbox{\includegraphics[width=0.225\textwidth,trim=5cm 5cm
    5cm 5cm,clip]{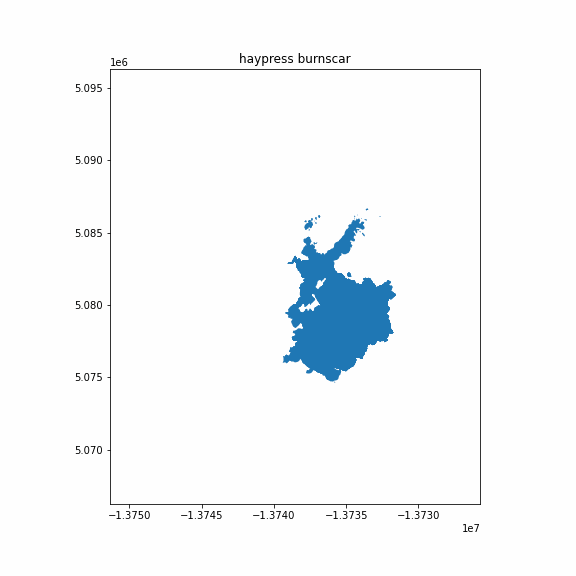}}
    \fbox{\includegraphics[width=0.225\textwidth,trim=5cm 5cm
    5cm 5cm,clip]{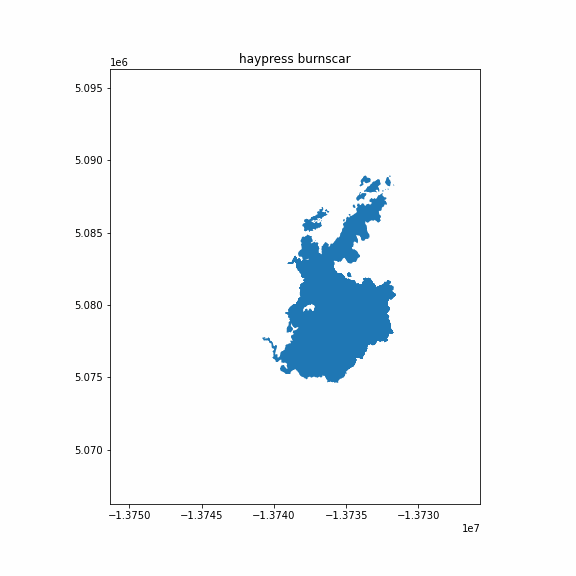}}
    \fbox{\includegraphics[width=0.225\textwidth,trim=5cm 5cm
    5cm 5cm,clip]{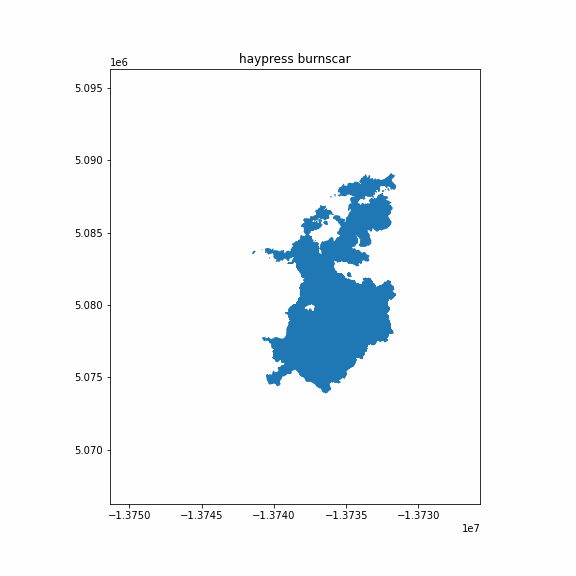}}
    \fbox{\includegraphics[width=0.225\textwidth,trim=5cm 5cm
    5cm 5cm,clip]{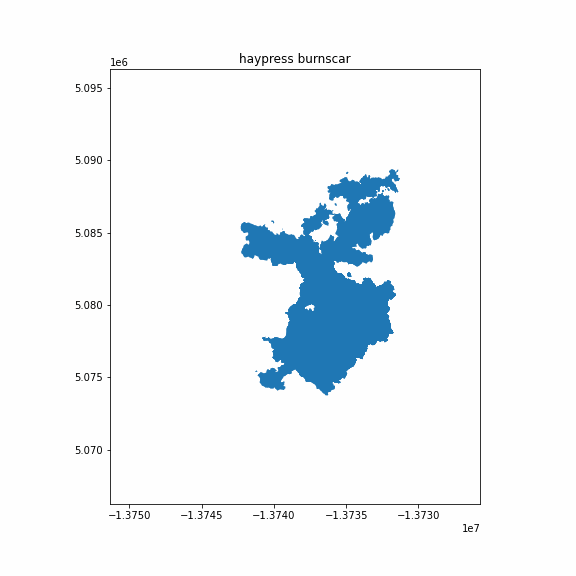}} \\[2pt]
    \fbox{\includegraphics[width=0.225\textwidth,trim=5cm 5cm
    5cm 5cm,clip]{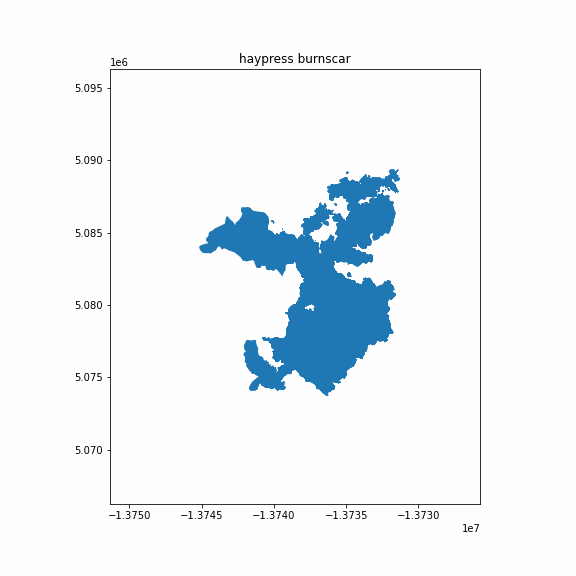}}
    \fbox{\includegraphics[width=0.225\textwidth,trim=5cm 5cm
    5cm 5cm,clip]{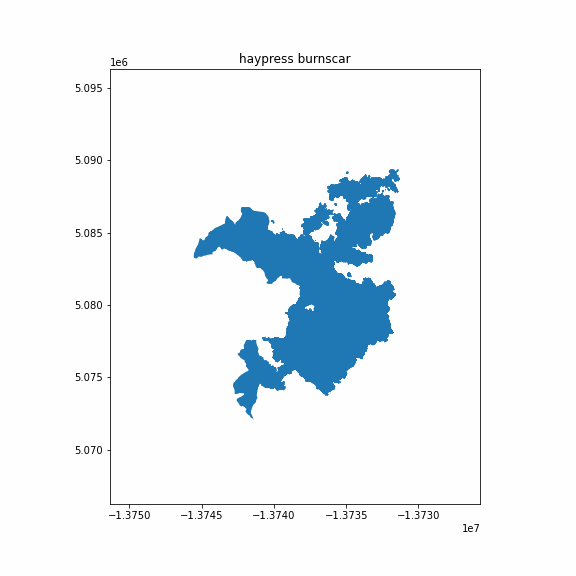}}
    \fbox{\includegraphics[width=0.225\textwidth,trim=5cm 5cm
    5cm 5cm,clip]{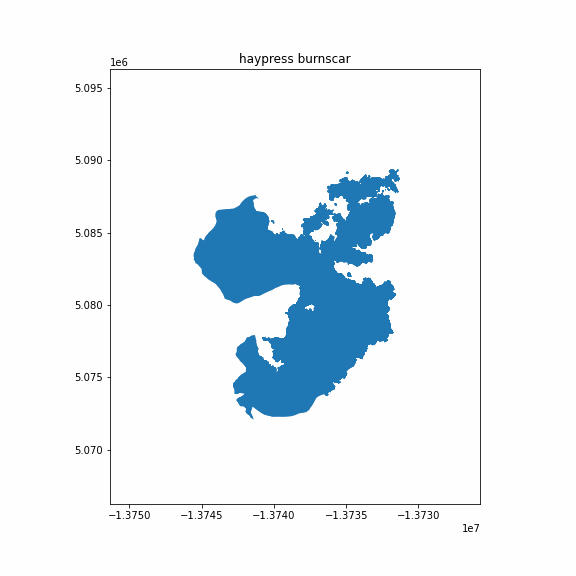}}
    \fbox{\includegraphics[width=0.225\textwidth,trim=5cm 5cm
    5cm 5cm,clip]{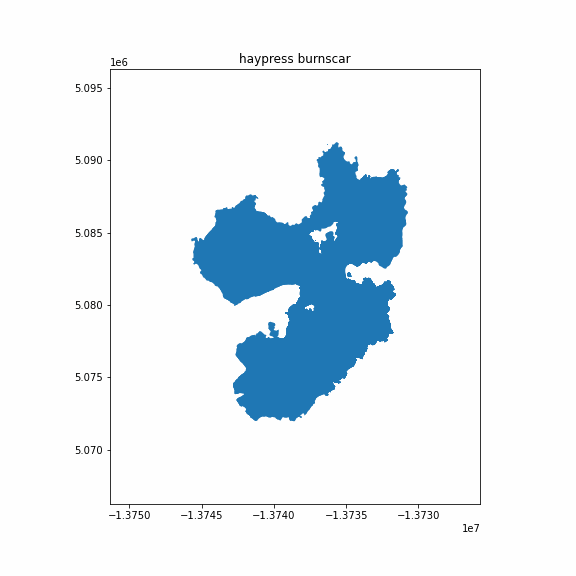}}
  \end{center}
  \caption{\label{fig:haypress} \em The Haypress fire exhibits the
  quadratic area scaling $A(t) \sim t^2$. Notice the abundance of
  spotting.}
\end{figure}

Figure~\ref{fig:historical2} shows that the area growth can also be
quadratic for other fires in the western USA. The black dashed lines
have slope 2, and the solid black line is an average over the seven fire
events. One fire event in Figure~\ref{fig:historical2} is the Haypress
fire, and GIS frames of the burn scar at different times are provided in
Figure~\ref{fig:haypress}. Importantly, the fire growth is clearly not
regular or isotropic, and isolated neighboring fires are evident,
indicating that another process is responsible for the quadratic growth.
We expect, and will show in subsequent sections, that nonlocal ember
spotting, which is observable in these frames, can result in quadratic
growth.

Similar growth rates in fire area have also been observed for western
USA fires between 2001 and 2020 by Juang {\em et
al.}~\cite{jua-wil-aba-bal-hur-mor2022} [Figure 2]. They analyzed a
large number of fires to find a statistical relation between the
increase in the size of the fire, $\Delta A$, and the fire size $A$,
that is, $A'(t) = \beta A(t)$. So long as $\beta \neq 1$, this implies
that
\begin{align}
  A(t) \sim t^{\frac{1}{1-\beta}}.
\end{align}
Averaging over 2,352 measurements, they calculate an average value of
$\beta = 0.46$, suggesting that the average area scaling result is $A(t)
\sim t^{1.85}$, however, their results are broadly distributed in
$\beta$ such that any slope between linear and quadratic scaling is also
likely.

These observations raise a central modeling question: {\bf What physical
mechanisms control the transition between a linear and quadratic growth
in a fire's area?} While plume-driven long-range spotting has been
studied extensively, ember wash has not, despite copious anecdotal
evidence for ember transport near the surface. The GIS data suggest
ember generated effects, near and far from a fire front, lead to rapid
growth. In particular, by comparing Figures~\ref{fig:goodview}
and~\ref{fig:haypress}, much more spotting events can be seen in the
Haypress fire whose area grew quadratically with time. The GIS-derived
fire growth behavior in Figure~\ref{fig:historical2}, and observations
discussed by Juang {\em et al.}~\cite{jua-wil-aba-bal-hur-mor2022}
suggest that fire areas can grow quadratically with time, and this
asymptotic growth lasting as long as three weeks. Models such as
FARSITE~\cite{fin1998} can generate quadratic growth by assuming that
the fire spreads radially along fixed axes of an ellipse.
However, assuming the shape of a fire amounts to a drastic modelling
choice that in this case immediately disallows linear growth in the fire
area, which is observed for the GIS data reported in
Figure~\ref{fig:historical1}. Rather than using a model that
automatically results in quadratic area growth, we propose a physical
model and extend it to processes that we expect play a significant role
in the asymptotic area growth.

Spot fire analyses show diverse patterns of near and long distance
ignition~\cite{page2019analysis, storey2020analysis}. Localized but
highly intermittent modes of transport can create spot fires, fingers,
but also widen the fire front without producing clearly separated spot
fires. Intense near-surface ember transport dominated by repeated
deposition and resuspension undoubtedly plays an important role in
shaping fire growth in many wildland fire situations.


\section{Saltation-Based Transport Model for Ember Wash}
\label{sec:saltation}

%
Motivated by the scaling behaviors identified in Section~\ref{sec:GIS},
we develop an ember wash model that is then coupled with a simplified 2D
surface wind model. Again, our goal is to seek simplified physical local
and nonlocal spread processes that generate the linear and quadratic
area growth. We use a standard sediment transport model
framework~\cite{meyer1948formulas, van2020fully,
southard2006introduction} 
to introduce the representation of ember wash in our 2D surface flow
model. The transport of embers ahead of the fire near the ground,
following the sediment transport analogy, is composed of ballistic hops
or {\em saltation}, sliding and rolling or {\em surface creep}, as well
as transport of less dense embers maintained in suspension by turbulent
eddies. Saltation is characterized by trajectories that rise sharply as
particles are lifted and accelerated downwind subject to lift and drag
to land at a distance much greater than the rise height. Surface creep
is characterized by substantial motion in contact with the ground. 

Our purpose is not to develop a detailed transport model due to the lack
of validation data, but rather to examine the key factors developed in
the sediment transport community, and then adapt these to construct a
plausible transport model for ember wash suitable for implementation in
our simple 2D wind model. Key differences between sediment transport and
ember wash are the evolution of ember mass and flight characteristics
due to ember combustion, and the spatial distribution of ember sources
in the moving fire front and in fires caused by the embers. For the
simplified model here we require a representation of ember velocities
and trajectories in the surface layer.

The force lifting the fallen ember arises from the surface shear
velocity, which may be approximated as $u^*=ku_{z_0}$ where $u_{z_0}$ is
the wind some height $z_0$, say 2~m above the ground ($u_2$), and $k <
1$ is a constant coefficient (in the case of sand transport $k \sim 0.1$
for $z_0=2$~m). The maximum shear velocity is also normally close to the
rms surface wind velocity $U_{\mathrm{rms}}$. For near-surface winds of
about 10~m/s, $u^* \sim$ 1--10~m/s. Embers in saltation tend to leave
the ground with speeds initially similar to $u^*$, then accelerate to
speeds closer to the near-surface (or canopy) mean wind speed $U_w$.
Embers in surface creep tend to move at speeds similar to $u^*$. Actual
ember velocities will be a relatively broad distribution around these
numbers, with both larger and smaller values due to the variation of
background wind with height and fluctuations in time.


For a rough estimate of an ember excursion, an ember scooped off the
ground in saltation mode by $u^*$ moves a horizontal distance $X_e$
given by the wind velocity $U_\mathrm{w}$ multiplied by a flight time
$\tau$. The time may be scaled by the inertial height $H = u^{*2}/2g$
divided by the settling velocity producing
\begin{align}
  \label{eqn:emberVel}
  X_e \approx U_\mathrm{w} \frac{H}{W_e}.
\end{align}
A formulation of the steady ember settling velocity is 
\begin{align}
W_e \approx
  \left(\frac{4}{3}gd(\rho_e/\rho_a)/C_d\right)^{1/2},
\end{align}
where $d$ is ember diameter, $g$ is gravity, $\rho_e$ and $\rho_a$ are densities of air and embers, and $C_d$ is a drag coefficient accelerating the ember in the wind $U_w$, hence
\begin{align}
  X_e \approx U_\mathrm{w} \frac{u^{*2}}{2g} \left(\frac{4}{3}gd(\rho_e/\rho_a)/C_d\right)^{-1/2}.
\end{align}
A wide range of excursions is expected due to the variety of ember
characteristics, flow dynamics, complex surface roughness with
vegetation, structures, and canopy effects. Only some of these effects
are present in the simplified model. As key wind parameters, $U_\mathrm{w}$
and $U_{\mathrm{rms}}$ play a fundamental role in surface ember
transport, and as with many boundary layer flow quantities they can be
defined in various ways according to the application and needs of the
model.

The purpose here is to introduce a representative ember excursion and
velocity as a first step toward a statistical representation of ember
wash in our idealized model. Typical distributions of sediment transport
hops show an exponential-like ember velocity
distribution~\cite{fur-sch-sch-fat2016}. These considerations motivate a
distance-based statistical description of ember wash, in which the
probability that an ember continues to be transported decays with the
distance it travels. This survival-based formulation is introduced in
the next section.

\section{Ember Wash as a Survival Analysis}
\label{sec:survival}
The saltation framework introduced in Section~\ref{sec:saltation}
typically describes sediment transport in terms of net fluxes and characteristic velocities. However, for a statistical representation of fire spread, the quantity of primary interest is not the instantaneous flux of all
embers, but whether embers travel far enough, while still burning, to
ignite new fuel. Near the surface, ember motion is intermittent and
repeatedly interrupted by deposition, trapping, burnout, or failure to
resuspend. These termination mechanisms act stochastically along an
ember's path, motivating a probabilistic description of ember travel
distance rather than a deterministic transport law. We use survival
analysis to justify the statistical models that we use in our numerical
results described in Section~\ref{sec:discussion}.

Let $R$ be the random variable of the travel distance of an ember in an
ember storm. Then, 
\begin{align}
  S(r) = P(R > r),
\end{align}
is the probability that an ember travels farther than distance $r$, and
this is the ember's survival function. It is related to the cumulative
density function $F_R$ through the identity $S = 1 - F_R$. Next, define
the hazard of an ember ignition per unit distance 
\begin{align}
  \kappa(r) = \lim_{\Delta r \rightarrow 0} 
    \frac{P(r<R<r+\Delta r \mid R>r)}{\Delta r}.
\end{align}
Physically, $\kappa(r)$ is the instantaneous risk per unit distance that
an ember ceases to be transported due to extinction, trapping, or loss
of mobility. Taking the limit, we have
\begin{align}
  \kappa(r) = -\frac{1}{S(r)}\frac{dS}{dr},
\end{align}
resulting in a governing equation for the survival function
\begin{align}
  \frac{dS}{dr} = -\kappa(r) S(r), \quad S(0) = 1.
\end{align}
The survival function depends on how we model the hazard function
$\kappa$. If $\kappa$ is independent of distance, corresponding to the
assumption that each additional meter of travel carries the same
probability of ember termination, then the survival function is
exponential:
\begin{align}
  \label{eqn:exponential}
  S(r) = \exp(-\kappa r) \Rightarrow 
    R \sim \texttt{Exp}(\kappa).
\end{align}
If $\kappa$ depends on $r$, then
\begin{align}
  S(r) = \exp\left(-\int_{0}^r \kappa(s)\, ds \right),
\end{align}
and if $\kappa$ varies slowly with $r$, then $R$ can be approximated
with an exponential distribution.

Given the hazard function, we convert the per distance hazard ($\kappa$)
to a per time hazard ($\lambda$), which depends on the fuel type,
trapping of embers, and environmental conditions. We define
\begin{align}
  \lambda = \kappa U_{\mathrm{w}},
\end{align}
where $U_{\mathrm{w}}$ is the near-surface wind from the saltation model
(equation~\eqref{eqn:emberVel}). This conversion allows the
distance-based survival model to be coupled consistently to the
time-stepping fire spread model, while preserving the interpretation of
$\kappa$ as a spatial termination rate.

The total distance traveled can be viewed as the accumulation of many
small spatial increments. The key assumption for exponential behavior is
that each additional meter of travel carries the same probability of
``survival'': extinction, trapping, or burnout, independent of the
distance already traveled. This corresponds to a constant hazard per
unit distance, meaning risk accumulates additively. A constant spatial
hazard uniquely leads to the exponential survival
law~\eqref{eqn:exponential}, so the total distance $R$ is exponential.
In Section~\ref{sec:fire_model}, this exponential survival model is
coupled to an idealized fire–atmosphere solver to examine how
near-surface ember wash modifies fire geometry, first-arrival times, and
area–time scaling.

\section{Coupling Ember Statistics to Fire Spread}
\label{sec:fire_model}
We couple the statistical model from Section~\ref{sec:survival} with an
idealized 2D model for near-surface wind in the presence of
fire~\cite{qua-spe2021}. For completeness, we briefly summarize the
original idealized model, with which we previously demonstrated various
effects of fire-atmosphere interactions, but in the absence of embers.
The model uses a 2D rasterized geometry, and each cell is always in one
of three states: actively burning; unburnt with available fuel; and,
burnt with no remaining fuel. Given a fire configuration, contributions
in the near-surface flow include a constant background wind, a divergent
flow induced by the buoyant fire plume, and a turbulent diffusion term.
The model for the divergent flow term was first introduced by Hilton
{\em et al.}~\cite{hil-sul-swe-sha-tho2018}, and they call this term the
pyrogenic potential.

Computing the flow field requires solving the two-dimensional Poisson
equation with a known (and evolving) forcing function. We use the
standard second-order central difference formula to discretize the
Laplacian, and we impose a constant Neumann boundary condition. The
resulting sparse linear system is solved with Matlab's backslash
operator. With the total velocity field in hand, new cells are ignited
downwind. We use Bresenham's line algorithm to carry the fire downwind.
Finally, burning cells are extinguished after they have combusted for a
user-specified amount of time, corresponding to fuel load.

In our setup, the cell size is 1~m$^{2}$, and the time evolution occurs
of minutes to tens of minutes. The full model domain is flat and, in
most runs, 200~m on each side. To minimize edge effects, we remove the
first 5~m of fuels around the entire perimeter of the domain. The only
fuel parameter in the model is a constant burn time which depends on the
fuel and combustion process and is regarded here as an observed
parameter. We use a value of 15~s which we estimated from
observations~\cite{cur-spe-hie-obr-goo-qua2018}. While the range of this
quantity can be large, mean values of a few seconds to minutes are
typical in light to moderate fuels. The strength of the pyrogenic source
is related to fire intensity by standard plume scaling.



Several new parameters are used to introduce ember washes into the
model. First, embers require a certain amount of shear to break free
from vegetation. Since our model calculates the near-surface horizontal
velocity at each cell in the grid, we define a threshold surface
velocity that must be exceeded for an ember to be created. Once an ember
is created, it travels with the near-surface velocity for a time $\tau$,
where $\tau$ is exponentially distributed with a mean value of $\mu$.
Finally, once the ember lands, a Bernoulli probability of ignition,
$\pig$, determines if the ember ignites the new cell.

\section{Results}
\label{sec:results}
%

\begin{figure}[htp]
  \centering
  \includegraphics[width=0.243\textwidth,trim=0.5cm 0.5cm 1.5cm
  1.0cm, clip=true]{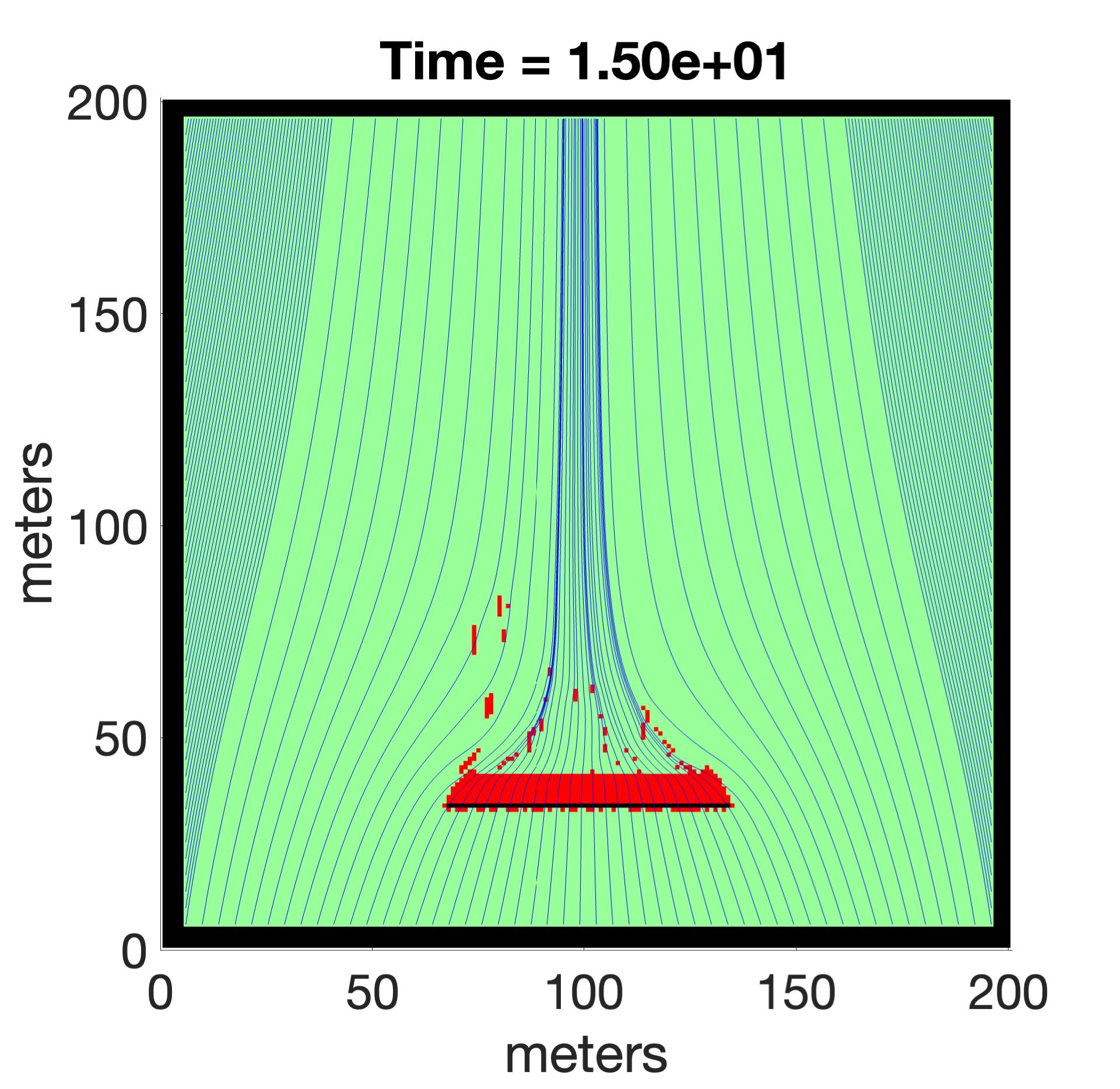}
  \includegraphics[width=0.243\textwidth,trim=0.5cm 0.5cm 1.5cm
  1.0cm, clip=true]{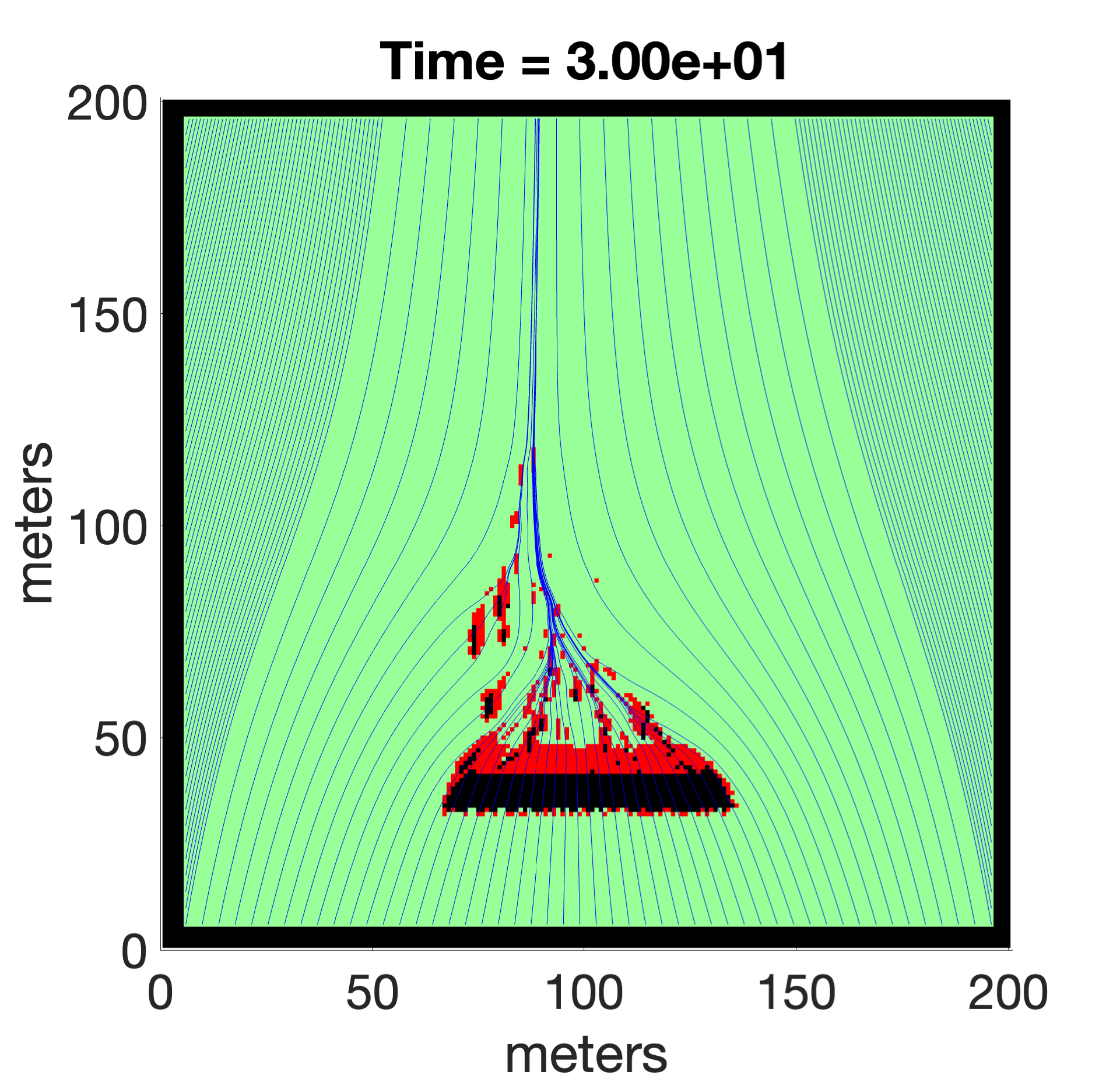}
  \includegraphics[width=0.243\textwidth,trim=0.5cm 0.5cm 1.5cm
  1.0cm, clip=true]{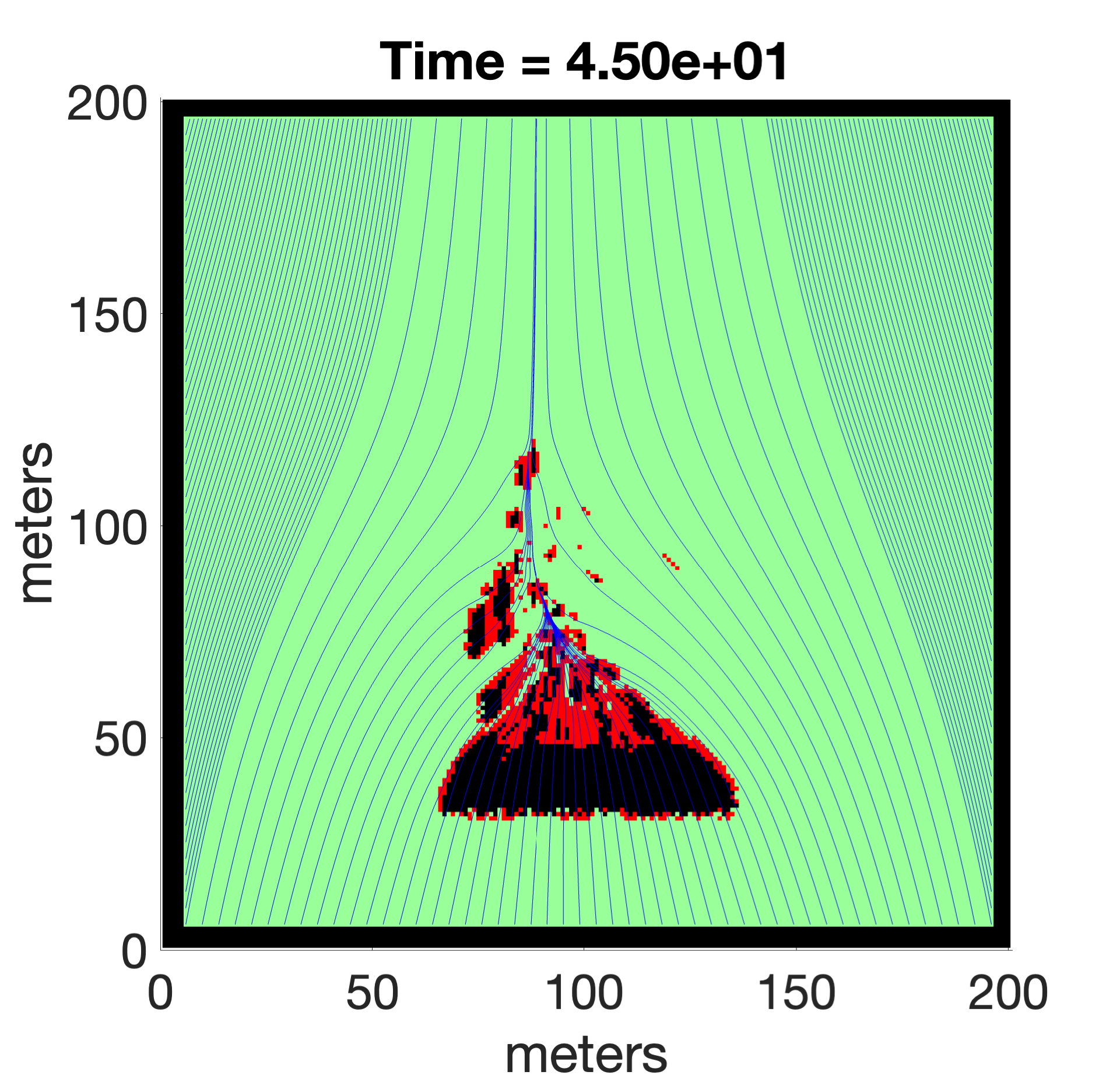}
  \includegraphics[width=0.243\textwidth,trim=0.5cm 0.5cm 1.5cm
  1.0cm, clip=true]{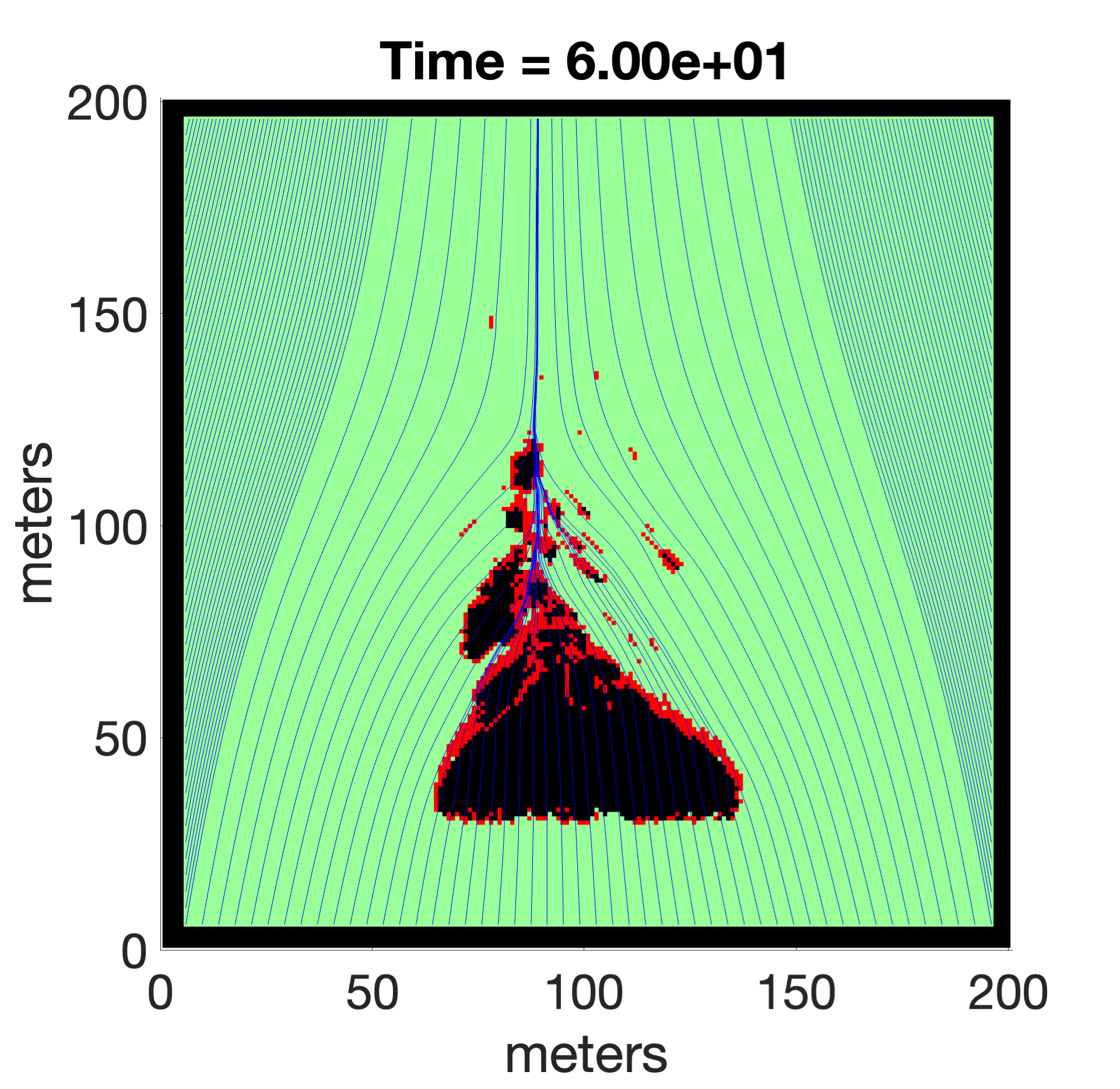}
  \\[10pt]
  \includegraphics[width=0.243\textwidth,trim=0.5cm 0.5cm 1.5cm
  1.0cm, clip=true]{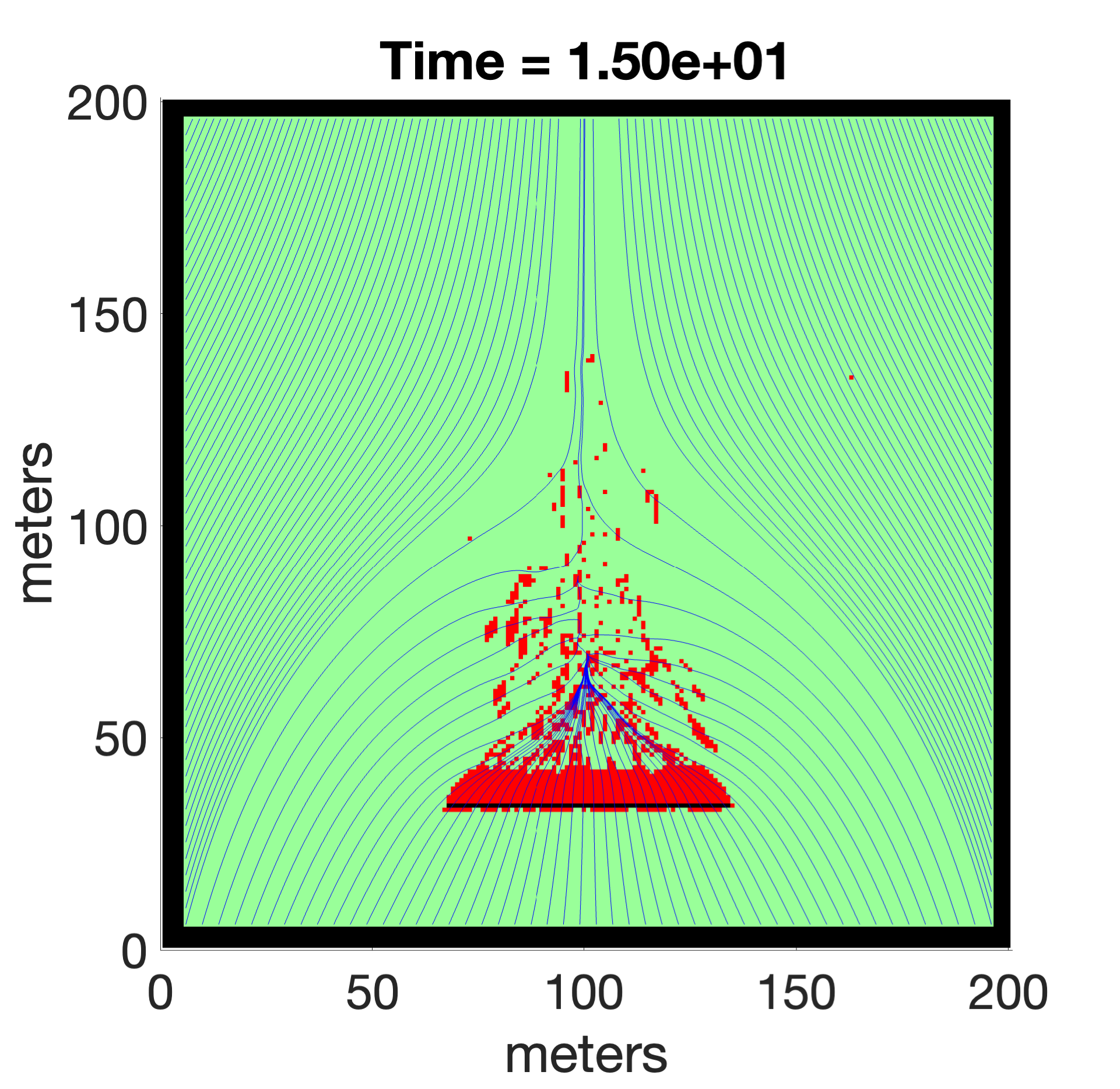}
  \includegraphics[width=0.243\textwidth,trim=0.5cm 0.5cm 1.5cm
  1.0cm, clip=true]{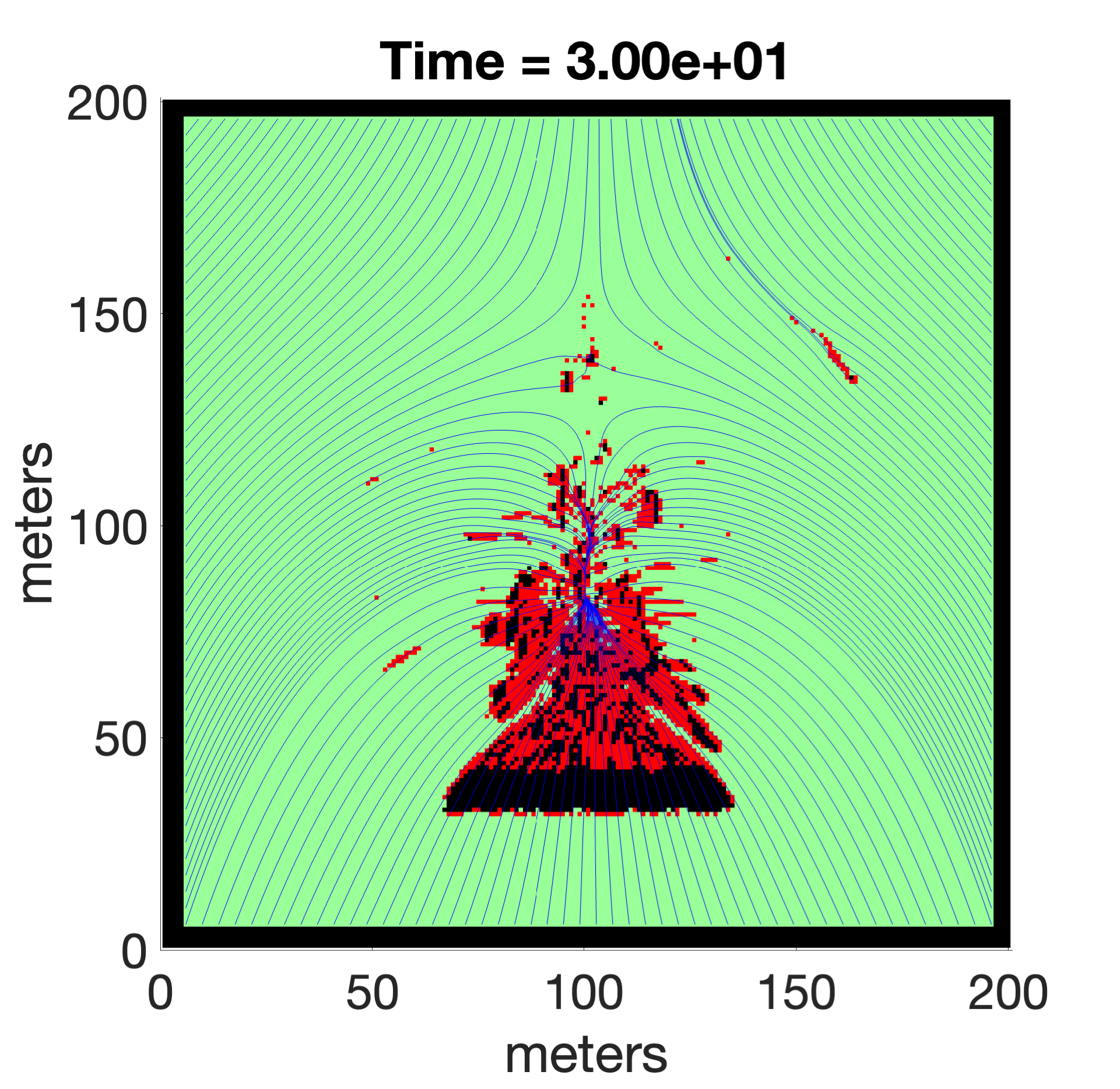}
  \includegraphics[width=0.243\textwidth,trim=0.5cm 0.5cm 1.5cm
  1.0cm, clip=true]{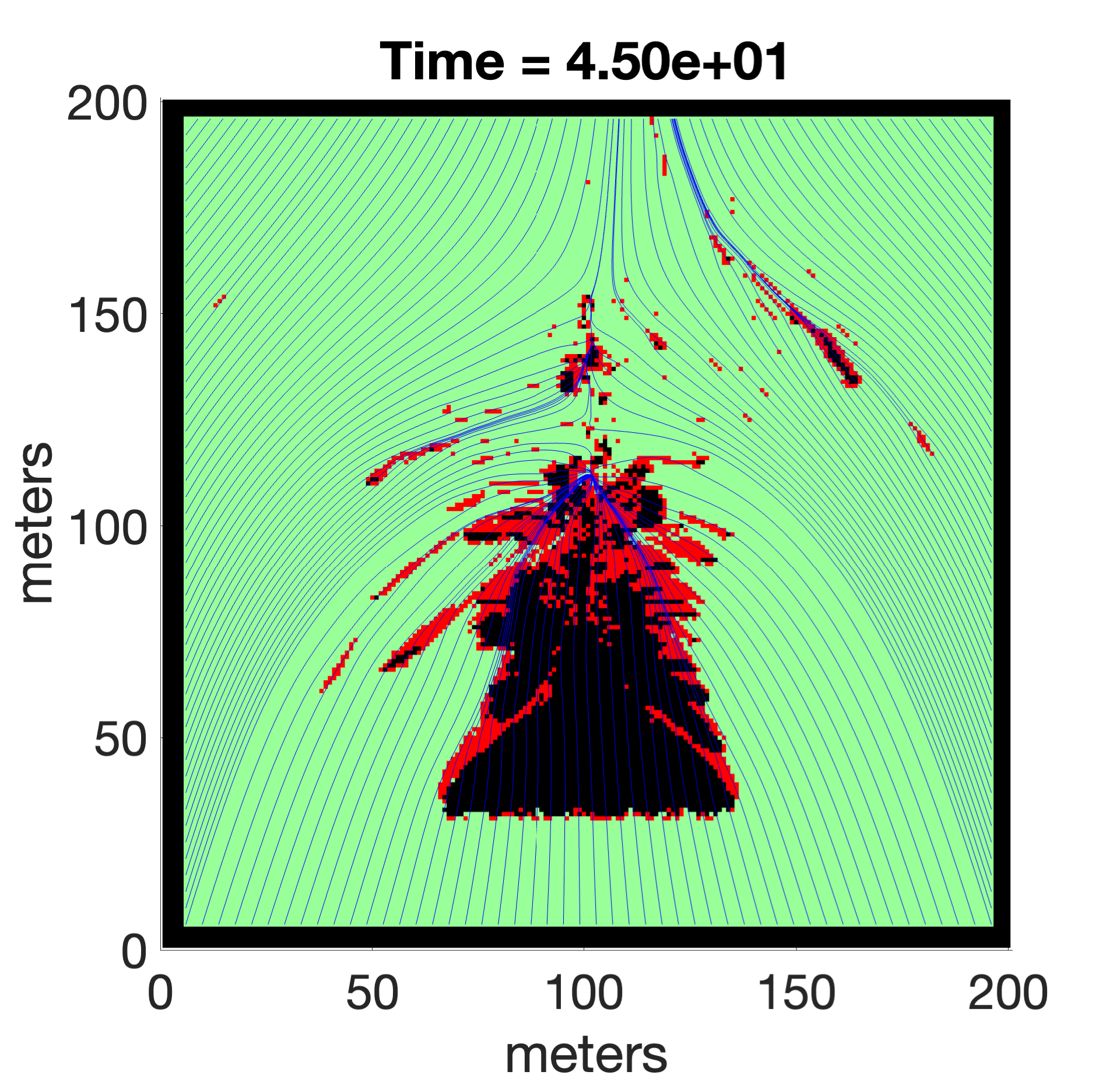}
  \includegraphics[width=0.243\textwidth,trim=0.5cm 0.5cm 1.5cm
  1.0cm, clip=true]{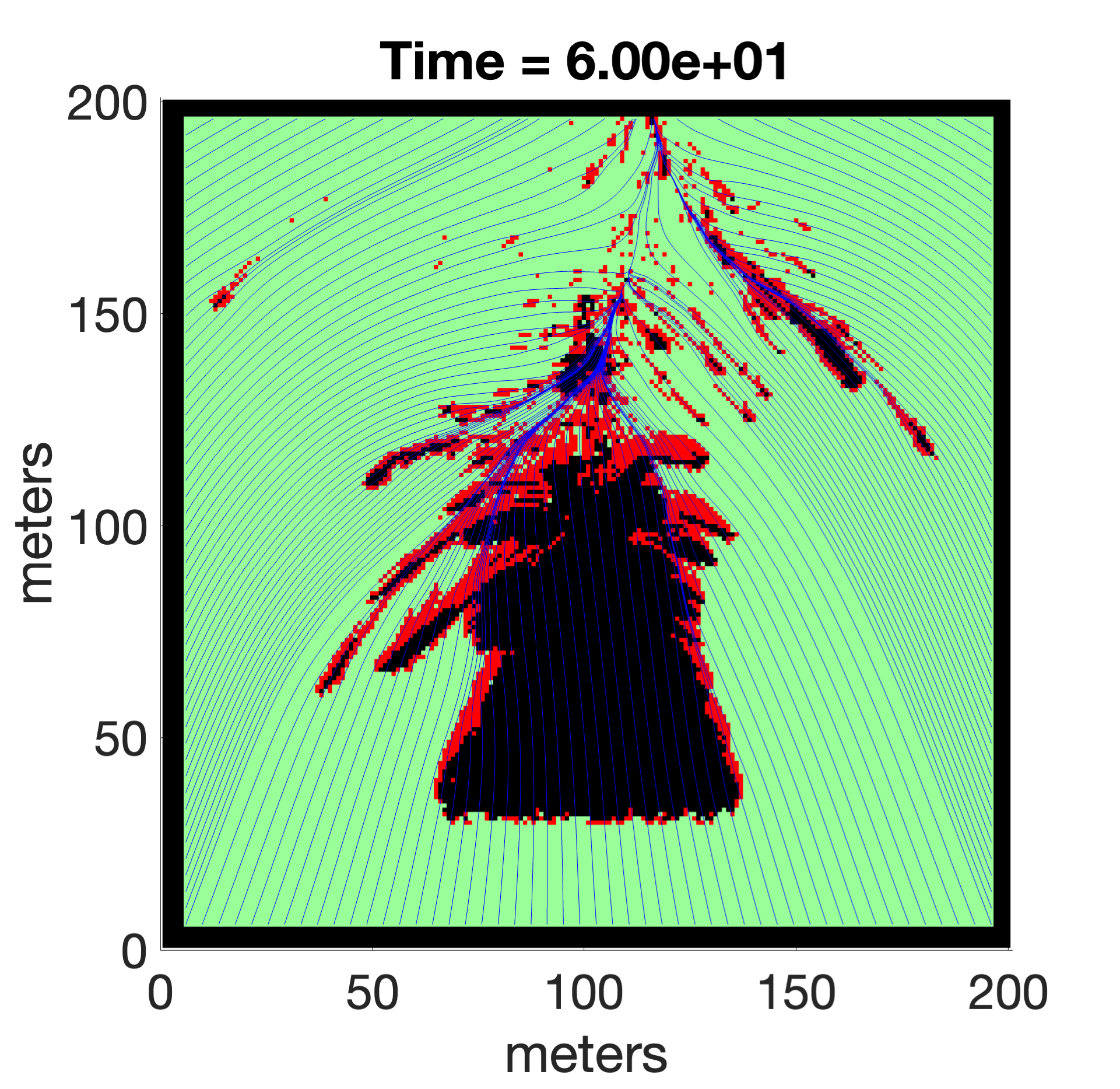}
  \caption{\label{fig:visualsegment} \em Idealized example of fire
  spread due to ember wash. The ember flight time is exponentially
  distributed with a mean of $\mu = 20~s$. The probability of ignition
  is $\pig = 0.1$ (top) and $\pig = 0.5$ (bottom). Streamlines of the
  wind (solid blue curves) indicate the interaction of a constant
  background wind of 1~m/s from the south with the fire-induced wind.}
\end{figure}

We have incorporated these parameters with the statistical ember model described in Section~\ref{sec:fire_model}. For all simulations, we use a threshold wind velocity value of 0.2~m/s. Given our model setup, smaller thresholds result in all combusting cells launching embers, while larger thresholds result in no ember generation. We vary the other two parameters, with exponential means 5~s $\leq \mu \leq$ 50~s and $\pig$ = 0.1 and 0.5. We considered intermediate probability of ignition values and find that the rate of area growth is smooth with respect to $\pig$.
Figure~\ref{fig:visualsegment} shows simulation results from the
simplified model coupled with the new ember model. The background wind
is 3~m/s, and the ember flight time is exponential distributed with a
mean value of $\mu = 20~s$. The probability of ignition is $\pig = 0.1$
(top) and $\pig = 0.5$ (bottom). The solid blue lines are streamlines of
the wind which affect the trajectory of embers.  As expected, the
presence of embers, especially for large probabilities of ignition,
results in a faster spread.

We non-dimensionalize the burn scar area by dividing $A(t)$ by the total
amount of available fuel. Instead of introducing a new variable, we
simply understand that $A(t) \in [0,1]$, with $A(t) = 1$ meaning that
all fuel has been consumed.

\begin{figure}[htp]
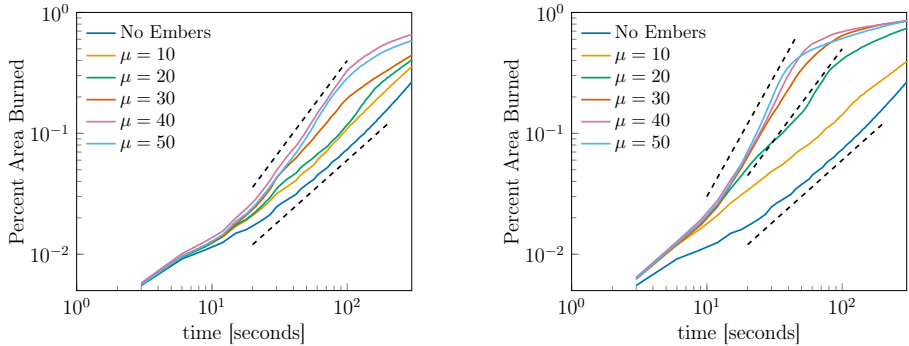

  \begin{center}
  \begin{tabular}{cc}
    \scalebox{0.65}{\input{area_vs_time_exponential_PIG0p1.tikz}} &
    \scalebox{0.65}{\input{area_vs_time_exponential_PIG0p5.tikz}}
  \end{tabular}
  \end{center}
  \caption{\label{fig:exponential1} \em The percentage of the available
  fuels that are actively burning or have been burnt. The probability of
  ignition is $\pig = 0.1$ (left) and $\pig = 0.5$ (right). The flight
  time of the embers follows an exponential distribution with the mean
  value provided in the legends. The black broken lines of slopes 1 and
  1.5 (left), and 1, 1.5, and 2 (right).}
\end{figure}

Figure~\ref{fig:exponential1} shows the percent area burned of a line
ignition parallel to the wind direction with $\pig = 0.1$ (left) and
$\pig = 0.5$ (right). Each colored line represents a different mean
flight time for embers, and snapshots of the simulations for $\mu =
20$~s are shown in Figure~\ref{fig:visualsegment}. The dashed black
lines have slopes 1, 1.5, and 2 and can be used to show how the rate of
area growth depends on $\mu$ and $\pig$. For the smaller probability of
ignition, we see a clear transition in the asymptotic scaling of the
percent area burned---without the ember storm, it grows linearly, and as
$\mu$ increases, it transitions to the scaling $A(t) \sim t^{1.5}$. With
the larger probability of ignition $\pig = 0.5$, the area growth covers
all regimes and scales all the way to $A(t) \sim t^{2}$.

\section{Discussion}
\label{sec:discussion}
The spatial patterns observed in the simulations provide insight into
the physical mechanisms governing ember transport and ignition.
Long-range spotting is commonly modeled as a stochastic transport
process in which firebrands are lofted into the atmosphere and then
advected by background winds before landing at some distance from the
source. Probabilistic models for landing distances have been developed
based on Lagrangian particle transport and turbulent
diffusion~\cite{sar-con-kai-fer-por2008}. In these models the primary
displacement mechanism is atmospheric advection well above the surface, producing landing distributions that are strongly biased in the background downwind direction.

The transport mechanism considered here is fundamentally different.
Ember wash occurs near the surface and is driven by turbulent structures
in the surface boundary layer rather than plume lofting. The resulting
transport is therefore more random and resembles saltation-like
transport. Consequently, ember wash also tends to broaden the fire
perimeter laterally, whereas plume spotting primarily accelerates
forward spread along the wind direction.

This distinction helps explain the spatial patterns observed in our
simulations. 
In our simulations the wind field is uniform and directed in a single
direction, so spotting produces an elongated ignition region extending
downwind of the original burn scar. In contrast, ember wash as defined
here spreads near the surface rather than aloft. Much of this spreading
may occur in a tumbling mode of transport with embers and debris
released and falling, bouncing, and rolling along the ground and being
resuspended and carried rapidly forward again in wind gusts. This
turbulence-driven suspended component probably includes a mix of very
short-range spotting and resuspended embers that have fallen near the
fire front.

Because ember wash occurs near the surface and is governed by stochastic
surface transport processes rather than plume advection, the resulting
ember displacement statistics differ fundamentally from those associated
with long-range spotting. We represent the ground level process by a
highly localized probability of ignition with an exponential decay away
from the fire front. Such a shape may be justified as a subset of the
observed close-range, exponential, spotting distributions in analyses of
spotting distance~\cite{storey2021experiments, page2019analysis}. The
appearance of exponential spread or displacement of the front is likely
a general result of the randomness of ember motion. In the sediment
transport case, some question of the exact distributions for hopping
distance and travel times remain~\cite{ancey2015stochastic}, while
experimental evidence suggests exponential-like distributions in
velocity and Weibull in hop distances~\cite{fathel2015experimental}.

Fire spread in highly constrained experimental pine straw bed conditions
without ember effects suggests an exponential distribution of front
displacements~\cite{sag-spe-pok-qua2021}. Here, the random nature of
turbulence in the wind at small scales, both at the fuel surface and
within the fuel bed, and in the buoyant plume is at play in the
convection of the hot gases and associated diffusive spread of fire
close to the front~\cite{beb-oli-qua-sko-hei-spe2020}. We expect a
fundamental role for boundary layer turbulence in the ember wash
phenomenon with complex coupled interactions between surface shear
layers, buoyancy, and roughness elements.

The probability of ignition along the Bresenham ignition line plays
several roles. One of the roles is the ember combustion lifetime; not
all embers in motion are burning, and not all that are burning will
ignite the fuel bed. Another role is the ignition probability of the
fuel bed itself, since embers may gather on nonflammable surfaces or
other locations with no ignition, as well as on fuels easily ignited. A
third might be the time the ember spends at rest before resuspension,
burning all the while. The nonlinear nature of the interaction with
local winds makes the fire spread sensitive to this parameter.

Near-surface ember transport can play a large role in wildfire spread,
yet it is not represented explicitly in most standard fire spread
models. The ember wash effect produces a larger burning area and hence
greater fire-induced wind component, as well as broader fires hence an
area-dependent fire spread. An extended fire width is associated with
complex time-dependent fire behaviour, apparent even with highly
simplified dynamics~\cite{qua-spe2021}. The possible range of turbulent
surface fire spread dynamics and effects is vast and only a restricted
set of conditions is explored here to demonstrate the applicability of
our approach using a physically constrained stochastic process model.
Comparisons with experimental results are needed to determine suitable
explicit model parameterizations of this surface mode of ember
transport. Despite the limitations of the 2D surface spread model, it
represents a significant range of this complex behaviour. The addition
of structures is straightforward and is presently under investigation in
a similar framework, but ultimately fully 3D coupled fire-atmosphere
simulations with a parameterization of the surface mode of transport are
needed.

\subsection*{CRediT author statement}
\medskip
\noindent
Conceptualization, K.S.~and B.Q.; methodology, K.S.~and B.Q.; software,
K.S.~and B.Q.; validation, K.S.~and B.Q.; formal analysis, K.S.~and
B.Q.; investigation, K.S.~and B.Q.; resources, K.S.~and B.Q.; data
curation, K.S.~and B.Q.; writing--original draft preparation, K.S.~and
B.Q.; writing--review and editing, K.S.~and B.Q.; visualization,
K.S.~and B.Q.; supervision, K.S.~and B.Q.; project administration,
K.S.~and B.Q.; funding acquisition, K.S.~and B.Q. All authors have read
and agreed to the published version of the manuscript.


\subsection*{Acknowledgments}
\medskip
\noindent
The GIS results were first processed by Craig Anderson and analyzed by
Xin Tong. This research was supported by the U.S.~Department of Defense,
Strategic Environmental Research and Development Program under Award
Number RC20-1298; the National Science Foundation, Division of
Atmospheric and Geospace Sciences, under Award Number 2427321; and by
the Geophysical Fluid Dynamics Institute, Florida State University.


\begin{thebibliography}{10}
\expandafter\ifx\csname url\endcsname\relax
  \def\url#1{\texttt{#1}}\fi
\expandafter\ifx\csname urlprefix\endcsname\relax\def\urlprefix{URL }\fi
\expandafter\ifx\csname href\endcsname\relax
  \def\href#1#2{#2} \def\path#1{#1}\fi

\bibitem{coe-cam-mic-pat-rig-yed2013}
J.~L. Coen, M.~Cameron, J.~Michalakes, E.~G. Patton, P.~J. Riggan, K.~M.
  Yedinak, {WRF-Fire: Coupled Weather–Wildland Fire Modeling with the Weather
  Research and Forecasting Model}, Journal of Applied Meteorology and
  Climatology 52 (2013) 16--38.

\bibitem{beb-oli-qua-sko-hei-spe2020}
Y.~Bebieva, J.~Oliveto, B.~Quaife, N.~Skowronski, W.~E. Heilman, K.~Speer, Role
  of horizontal eddy diffusivity within the canopy on fire spread, Atmosphere
  11~(6) (2020) 672.

\bibitem{albini2012mathematical}
F.~A. Albini, M.~E. Alexander, M.~G. Cruz, A mathematical model for predicting
  the maximum potential spotting distance from a crown fire, International
  Journal of Wildland Fire 21~(5) (2012) 609--627.

\bibitem{martin2016spotting}
J.~Martin, T.~Hillen, The spotting distribution of wildfires, Applied Sciences
  6~(6) (2016) 177.

\bibitem{bhutia2010comparison}
S.~Bhutia, M.~Ann~Jenkins, R.~Sun, Comparison of firebrand propagation
  prediction by a plume model and a coupled--fire/atmosphere large--eddy
  simulator, Journal of Advances in Modeling Earth Systems 2~(1) (2010).

\bibitem{storey2021experiments}
M.~A. Storey, O.~F. Price, M.~Almeida, C.~Ribeiro, R.~A. Bradstock, J.~J.
  Sharples, Experiments on the influence of spot fire and topography
  interaction on fire rate of spread, Plos one 16~(1) (2021) e0245132.

\bibitem{page2019analysis}
W.~G. Page, N.~S. Wagenbrenner, B.~W. Butler, D.~L. Blunck, {An analysis of
  spotting distances during the 2017 fire season in the Northern Rockies, USA},
  Canadian Journal of Forest Research 49~(3) (2019) 317--325.

\bibitem{sardoy2007modeling}
N.~Sardoy, J.-L. Consalvi, B.~Porterie, A.~C. Fernandez-Pello, Modeling
  transport and combustion of firebrands from burning trees, Combustion and
  Flame 150~(3) (2007) 151--169.

\bibitem{tohidi2017stochastic}
A.~Tohidi, N.~B. Kaye, Stochastic modeling of firebrand shower scenarios, Fire
  Safety Journal 91 (2017) 91--102.

\bibitem{manzello2020role}
S.~L. Manzello, S.~Suzuki, M.~J. Gollner, A.~C. Fernandez-Pello, Role of
  firebrand combustion in large outdoor fire spread, Progress in energy and
  combustion science 76 (2020) 100801.

\bibitem{dos-yag2023}
I.~D.-R. dos Santos, N.~Yaghoobian, {Effects of Urban Boundary Layer Turbulence
  on Firebrand Transport}, Fire Safety Journal 135 (2023) 103726.

\bibitem{wad-sut-ooi-moi2022}
R.~Wadhwani, D.~Sutherland, A.~Ooi, K.~Moinuddin, {Firebrand transport from a
  novel firebrand generator: Numerical simulation of laboratory experiments},
  International Journal of Wildland Fire 31~(6) (2022) 634--648.

\bibitem{koo-pag-wei-woy2010}
E.~Koo, P.~Pagni, D.~Weise, J.~P. Woycheese, Firebrands and spotting ignition
  in large-scale fires, International Journal of Wildland Fire 19 (2010)
  818--843.

\bibitem{des-goo-ban2022}
A.~Desai, S.~Goodrick, T.~Banerjee, Investigating the turbulent dynamics of
  small‑scale surface fires, Scientific Reports 12 (2022) 10503.

\bibitem{fur-sch-sch-fat2016}
D.~J. Furbish, M.~W. Schmeeckle, R.~Schumer, S.~L. Fathel, {Probability
  distributions of bed load particle velocities, accelerations, hop distances,
  and travel times informed by Jaynes’s principle of maximum entropy},
  Journal of Geophysical Research: Earth Surface 121 (2016) 1373--1390.

\bibitem{jua-wil-aba-bal-hur-mor2022}
C.~S. Juang, A.~P. Williams, J.~T. Abatzoglou, J.~Balch, M.~D. Hurteau,
  M.~Mortiz, {Rapid Growth of Large Forest Fires Drives the Exponential
  Response of Annual Forest-Fire Area to Aridity in the Western United States},
  Geophysical Research Letters 49 (2022) e2021GL097131.

\bibitem{fin1998}
M.~A. Finney, {FARSITE, Fire Area Simulator--model development and evaluation},
  Vol.~3, US Department of Agriculture, Forest Service, Rocky Mountain Research
  Station, 1998.

\bibitem{storey2020analysis}
M.~A. Storey, O.~F. Price, R.~A. Bradstock, J.~J. Sharples, {Analysis of
  variation in distance, number, and distribution of spotting in southeast
  Australian wildfires}, Fire 3~(2) (2020) 10.

\bibitem{meyer1948formulas}
E.~Meyer-Peter, R.~M{\"u}ller, Formulas for bed-load transport, in: IAHSR 2nd
  meeting, Stockholm, appendix 2, IAHR, 1948.

\bibitem{van2020fully}
L.~Van~Rijn, G.~Strypsteen, A fully predictive model for aeolian sand
  transport, Coastal Engineering 156 (2020) 103600.

\bibitem{southard2006introduction}
J.~Southard, Introduction to fluid motions, sediment transport and
  current-generated sedimentary structures, 2006.

\bibitem{qua-spe2021}
B.~Quaife, K.~Speer, {A Simple Model for Wildland Fire Vortex–Sink
  Interactions}, Atmosphere 12 (2021) 1014.

\bibitem{hil-sul-swe-sha-tho2018}
J.~Hilton, A.~Sullivan, W.~S.~J. Sharples, C.~Thomas, Incorporating convective
  feedback in wildfire simulations using pyrogenic potential, Environmental
  Modelling \& Software 107 (2018) 12--24.

\bibitem{cur-spe-hie-obr-goo-qua2018}
M.~Currie, K.~Speer, K.~Hiers, J.~O'Brien, S.~Goodrick, B.~Quaife, {Pixel-Level
  Statistical Analyses of Prescribed Fire Spread}, Canadian Journal of Forest
  Research 49~(1) (2018) 18--26.

\bibitem{sar-con-kai-fer-por2008}
N.~Sardoy, J.~Consalvi, A.~Kaiss, A.~Fernandez-Pello, B.~Porterie, Numerical
  study of ground-level distribution of firebrands generated by line fires,
  Combustion and Flame 154 (2008) 478--488.

\bibitem{ancey2015stochastic}
C.~Ancey, P.~Bohorquez, J.~Heyman, Stochastic interpretation of the
  advection-diffusion equation and its relevance to bed load transport, Journal
  of Geophysical Research: Earth Surface 120~(12) (2015) 2529--2551.

\bibitem{fathel2015experimental}
S.~L. Fathel, D.~J. Furbish, M.~W. Schmeeckle, Experimental evidence of
  statistical ensemble behavior in bed load sediment transport, Journal of
  Geophysical Research: Earth Surface 120~(11) (2015) 2298--2317.

\bibitem{sag-spe-pok-qua2021}
D.~Sagel, K.~Speer, S.~Pokswinski, B.~Quaife, {Fine-Scale Fire Spread in Pine
  Straw}, Fire 4 (2021) 69.

\end{thebibliography}
\end{document}